\documentclass[journal]{IEEEtran}

\ifCLASSINFOpdf

\else

\fi

\usepackage{listings}
\usepackage{multirow}
\usepackage[table,xcdraw]{xcolor}
\usepackage[table]{xcolor}
\usepackage{pifont}
\newcommand{\cmark}{\ding{52}}%
\newcommand{\xmark}{\ding{56}}%
\usepackage{amsmath}
\usepackage{makecell}
\usepackage[linesnumbered, ruled, vlined]{algorithm2e}
\usepackage{algpseudocode}
\usepackage{float}
\usepackage{graphicx}
\usepackage{color,soul}
\usepackage{cite}
\usepackage[normalem]{ulem}
\usepackage[bookmarks=false,hidelinks]{hyperref}
\renewcommand\hl[1]{#1}

\begin{document}

\title{ZigZag: A Memory-Centric Rapid DNN Accelerator Design Space Exploration Framework}

\author{Linyan~Mei$^\dagger$, 
Pouya~Houshmand$^\dagger$, Vikram~Jain,
Sebastian~Giraldo,~and~Marian~Verhelst% <-this % stops a space
\\
\vspace{0.25cm}
MICAS, ESAT, KU Leuven
\\
\vspace{0.3cm}
\footnotesize{$^\dagger$These~authors~contributed~equally~to~this~work.}
% \vspace{-0.1cm}\thanks{linyan.mei@kuleuven.be} 
%\thanks{Contact: \{linyan.mei, pouya.houshmand\}@kuleuven.be}
% \thanks{* We will put ZigZig open-source soon.}
% \thanks{Digital Object Identifier ...}
}

% make the title area
\maketitle

% As a general rule, do not put math, special symbols or citations
% in the abstract or keywords.
\begin{abstract}
Building efficient embedded deep learning systems requires a tight co-design between DNN algorithms, memory hierarchy, and dataflow. However, owing to the large degrees of freedom in the design space, finding an optimal solution through the implementation of individual design points becomes infeasible. Recently, several estimation frameworks for fast design space exploration (DSE) have emerged, yet they either suffer from long runtimes or a limited exploration space. This work introduces ZigZag, a memory-centric rapid DNN accelerator DSE framework which extends the DSE with uneven mapping opportunities, in which operands at shared memory levels are no longer bound to use the same memory levels for each loop index. For this, ZigZag uses a memory-centric nested-for-loop format as a uniform representation to integrate algorithm, accelerator, and algorithm-to-accelerator mapping, and consists of three key components: 1) a latency-enhanced analytical Hardware Cost Estimator, 2) a Temporal Mapping Generator that supports even/uneven scheduling on any type of memory hierarchy, and 3) an Architecture Generator that explores the whole memory hierarchy design space. Benchmarking experiments against existing frameworks, together with three case studies at different design abstraction levels show the strength of ZigZag. Up to 33\% more energy-efficient solutions are found by introducing ZigZag's uneven scheduling opportunities.
\end{abstract}

% Note that keywords are not normally used for peerreview papers.
\begin{IEEEkeywords}
Deep neural networks, accelerator, cost model, dataflow, mapping, scheduling, design space exploration.

\vspace{1.34ex}
{\href{https://github.com/ZigZag-Project/zigzag}{\bfseries\textit{Source~code}---Source code is available at https://github.com/\nolinebreak[0]ZigZag-Project/zigzag.}}

\end{IEEEkeywords}

\IEEEpeerreviewmaketitle

% ----\input{1_intro}------

\section{Introduction}
\label{section_intro}

Over the last decade, deep neural networks (DNNs) have established themselves as the principal algorithm for pattern recognition and data mining tasks, dominating the field of artificial intelligence (AI). %Subsequent to their success, the number of DNN models and their applications has seen an exponential growth, \mv{recently also} for embedded systems. 
Recent DNN models achieve greatly improved accuracies at the expense of increased depth and complexity. Executing these complex models on embedded systems becomes challenging due to resource and power constraints in edge devices. 

To meet the constraints in these devices, a lot of research in the recent past, in both industry and academia, has been done towards developing specialized hardware accelerators ~\cite{ENVISION,EyerissV1, binareye, EIE, laika, DNPU, Fused, diannao, shidiannao, TPU, scnn} for energy-efficient and high-throughput mapping of DNN workloads. Each accelerator is designed with a different memory hierarchy and a different choice of dataflow.
%The throughput and energy-efficiency of a given dataflow not only depends on the DNN workload (i.e. layer shape and size) but also the hardware architecture (memory hierarchy, precision, bandwidth). 
However, most of them are ad-hoc and local optimal designs resulting from exploration of a limited design space. It is hard to say if the configuration selected by the accelerators is the best one, given the vast design space available. Therefore it is essential to have a framework that can rapidly explore the available design space to guide designers in finding the Pareto optimal architectures with the optimal dataflow while taking in the hardware constraints and algorithmic workloads.

\begin{figure}[t]
\centering
\includegraphics[width=3.3in]{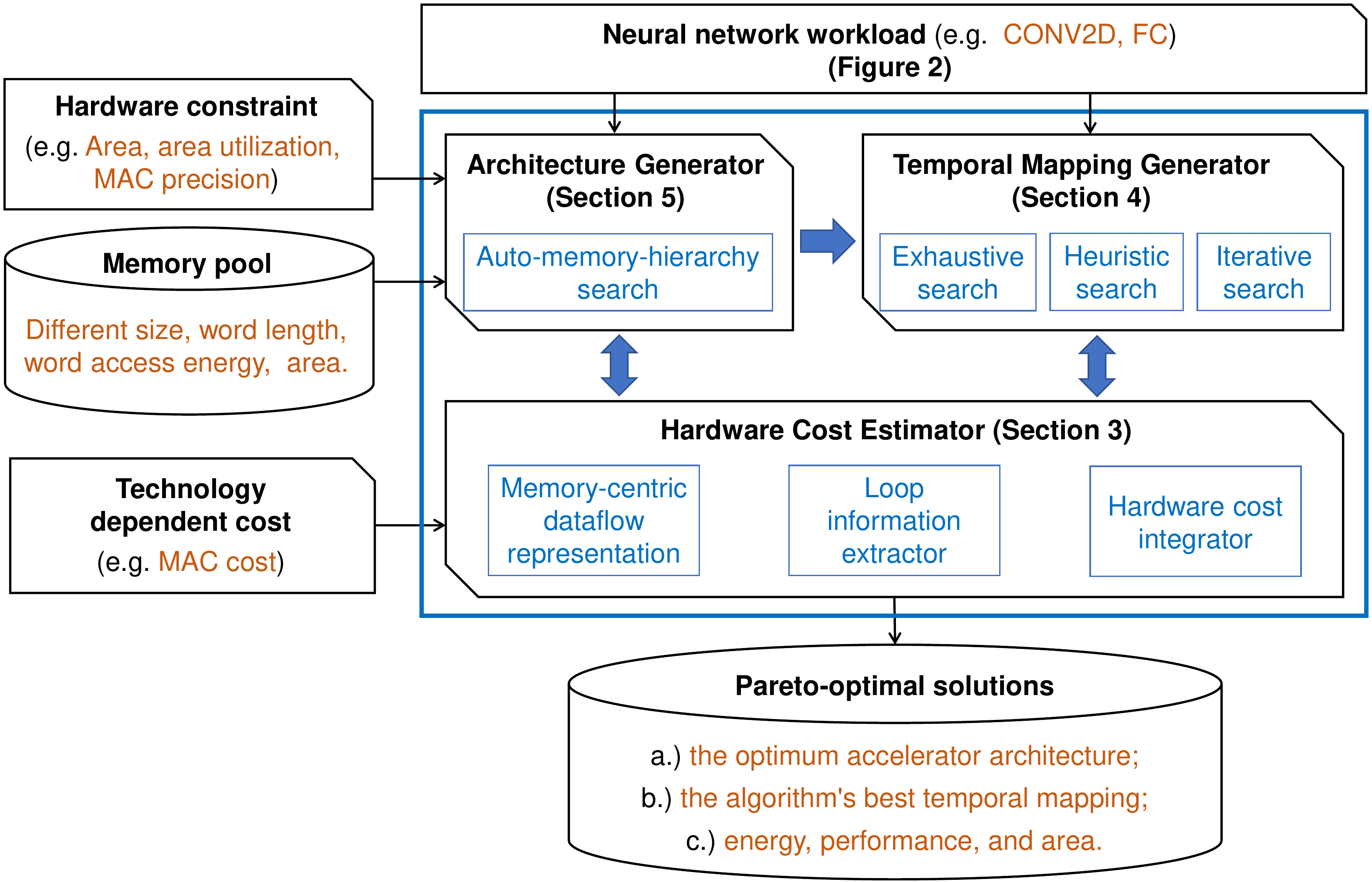}
% \vspace{-1.2em}
\caption{\hl{ZigZag framework diagram.}}
% \vspace{-1em}
\label{fig:framework_diagram}
\end{figure}

Many frameworks have thus emerged over the last few years targeting such hardware-software co-optimization by exploring the large design space available in the DNN system. Recent works on DSE framework in literature include 
%, which are also used as benchmarks for this work, \mv{I would not say this yet.}
Interstellar \cite{Yang2018}, SMAUG \cite{SMAUG}, Accelergy \cite{Accelergy}, Dory \cite{Dory}, Timeloop \cite{TIMELOOP}, dMazeRunner \cite{DMAZERUNNER}, MAESTRO \cite{MAESTRO}, and MAGnet \cite{MAGNET}. %\mv{We miss Accelergy?}
%\mv{I am not a fan for squeezing the SotA discussion into the introduction for 4 reasons: 1.) it makes your SotA discussion a bit invisible, and if reviewers will be asked whether you survey the SotA well, they might say no, because it is somewhat hidden; 2.) your introduction becomes very long. It is best to finish the introduction in a single page; 3.) I am very sad that your SotA figure disappeared. I really loved it and it clearly showed our ambition. I assume this is due to lack of space, but can we really not find a way to put it back? Now it is hard to understand your "classification" by only reading about it, and not having a way to visually picture it... 4.) it is always a good idea to have a (small) figure on page 1: this sticks with the reviewers, and they will remember the paper better. The SotA figure was ideal for this.} 
In order to provide a clear view in this domain, Table I compares different DSE frameworks' design space, search engine, and cost estimation strategy.

\begin{table*}[h!]
\centering
\caption{\hl{DNN accelerator DSE framework comparison}}
\label{fig:framework_comparison_table}
\scalebox{0.98}{
\begin{tabular}{|l|l|l|l|l|l|}
\hline
\textbf{Framework} & \textbf{Hardware design space} & \makecell{\textbf{Temporal mapping} \\ \textbf{space}} & \makecell{\textbf{Temporal mapping} \\ \textbf{generator}} & \textbf{Cost estimation}\\
\hline

Timeloop+Accelergy\cite{TIMELOOP,Accelergy} & {Fully flexible} & Even mappings & Constraint-driven & Highly fine-grained analy. model \\
\hline
MAESTRO\cite{MAESTRO} & Lowly flexible HW template & Even mappings & \makecell[l]{Predefined IS/WS/OS/RS} & Coarse-grained analytical model \\
\hline
\makecell[l]{Interstellar\cite{Yang2018}\\} & All mem. levels shared & Even mappings & Fully flexible & Coarse-grained analytical model \\
\hline
dMazeRunner\cite{DMAZERUNNER} & All mem. levels shared & Even mappings & Constraint-driven & Coarse-grained analytical model \\
\hline
MAGnet\cite{MAGNET} & Highly flexible HW template & Even mappings & Constraint-driven & \makecell[l]{Real hardware implementation} \\
\hline
Dory\cite{Dory} & Fixed architecture & Even mappings & Optimization-based & Coarse-grained analytical model \\
\hline
SMAUG\cite{SMAUG} & Fixed architecture & Even mappings & Fixed mapping & Cycle-accurate estimator \\
\hline
Ours: ZigZag & Fully flexible & \makecell[l]{Even and uneven \\ mappings} & \makecell[l]{Fully flexible with \\ optional constraints} & Fine-grained analytical model \\
\hline
\end{tabular}
}
% \vspace{-1em}
\end{table*}

\hl{Firstly, there are three main design spaces that need to be considered in the DSE, which are the algorithm space, hardware space, and algorithm-to-hardware mapping space. Most DNN-accelerator-DSE frameworks focus on exploring the latter two. Concerning the hardware design space, some SotA frameworks support a fully flexible hardware configuration (PE array and memory hierarchy), like Timeloop}~\cite{TIMELOOP}; while others pre-define a hardware template with certain tunable parameters, like MAGnet~\cite{MAGNET}; a last group makes specific assumptions, such as the sharing of all memory levels for the operands I/W/O and only explore within these constraints, like Insterstellar~\cite{Yang2018}. Concerning the temporal mapping space (scheduling space), all of the SotA frameworks only support even mappings, i.e. different operands need to follow the same loop blocking scheme at each shared memory level. Even and uneven mapping/loop blocking will be discussed in more detail further in this paper.

Secondly, the DSE tools typically encompass a temporal mapping generator, a.k.a. auto-scheduler or mapper, to find the optimum (in energy or/and performance) temporal mapping scheme for mapping a certain neural network layer onto a certain accelerator architecture. Most of the DSE frameworks perform a constraint-driven search to narrow down the space and speed up the searching procedure, like deMazeRunner~\cite{DMAZERUNNER}; some formulate the scheduling process into a linear problem and utilize optimization tools to solve it, like Dory~\cite{Dory}; the other use predefined dataflow to generate valid mapping points, like MAESTRO~\cite{MAESTRO}. Commonly used scheduling constraints and strategies include setting threshold for memory/PE array utilization and data reuse factor, and putting optimization goal for minimize certain cost functions, like the DRAM access or the overall memory traffic.

Finally, the last column in Table 1 listed the cost estimating approach adopted by each framework, in which three main categories can be noticed, 1) slow but very accurate hardware implementation based on High-Level Synthesis (HLS)~\cite{MAGNET}, 2) medium-speed and accurate cycle-accurate system simulator~\cite{SMAUG}, and 3) fast and generally accurate analytical model. Moreover, there are different granularity levels for analytical model~\cite{Accelergy}. Models, which distinguish memory writing from reading, consider memory word-width's impact on access cost, and take data pattern/stationarity into account in unit cost, are referred to as fine-grained models.

%\newpage    
This work proposes ZigZag, a memory-centric rapid DNN accelerator DSE framework (Section 2). 
%ZigZag brings the best of both worlds by using the faster analytical cost models and large design space exploration strategies to provide accurate estimations and locate optimum solution.  % belonging to the last category of frameworks. 
%ZigZag takes in deep learning workloads and hardware constraints as input, and produces many Pareto-optimal solutions as output. Each consists of the optimum memory hierarchy, the algorithm's best schedule, and the resulting execution latency, throughput, energy, and area.
Zigzag innovates on broadening the architecture and scheduling searching space, especially on enabling fully-flexible memory hierarchies search and even/uneven auto-scheduling, and thus discovers better design points than other frameworks.
%\lm{This is in comparison to the SotA DSE frameworks which either require a pre-defined memory hierarchy templates (Dory, SMAUG, MAESTRO, MAGnet) or can automatically generate memory hierarchies with whatever user-defined level numbers but have to follow some strict constraints like all memory levels are shared between operands W/I/O (Interstellar, dMazeRunner). Furthermore, none of the SotA frameworks supports uneven scheduling like ZigZag does, which opens a whole new space in DSE and discovers better design points.} 
ZigZag can also estimate performance (latency/throughput/MAC array utilization), important metrics of an accelerator that are lacking in some of the other frameworks. ZigZag estimates performance not only based on spatial mapping but also memory bandwidth and computing capacity. 
% Moreover, unlike some other frameworks which either uses exhaustive search but takes a long time or uses in-complete search that , 
Moreover, our framework uses smarter searching strategies to explore the enlarged design space, reducing the runtime while still locating the global optimum design point as exhaustive search does. To sum up, ZigZag made the following three key contributions. %\seb{comment: You can make bullet points for each contribution and maybe synthesize more the information to make the contribution more easy to grasp }
%\mv{I agree with comment of Seb that contributions could be shorter and/or bulletized, to make them stand out.}

%ZigZag's idea started from Yang et. al's framework~\cite{Yang2018} and then extended towards handling a boarder range of architectures, including memory hierarchies with levels shared among different operands, in a more accurate yet fast way. ZigZag makes the following four key contributions.

Firstly, the \textbf{memory-centric dataflow representation (Section 3)}, based on an enhanced nested-for-loop format, is proposed as a uniform representation for each design point. It integrates the information of algorithm, accelerator, and algorithm-to-accelerator spatial \& temporal mapping (a.k.a. dataflow). This newly-proposed representation opens up a whole new space for DSE  by decoupling the operands (W/I/O), the memory hierarchy, and the mapping scenarios. Combined with a proposed \textit{loop relevance principle}, the framework extracts in a systematic and insightful way the key information like number of memory accesses, required memory bandwidth, etc., to derive the system's energy and performance. 

%It uses 3 sets of loops to separately handle three DNN operands, input(I), filter weight(W), and output(O).

Secondly, the \textbf{Temporal Mapping Generator (Section 4)} is built to generate valid temporal mapping points for any type of memory hierarchy, in which each memory level for each operand can be shared or separated, with or without spatial unrolling, under even or uneven blocking. Additionally, to cope with the enlarged design space, three fast search strategies are proposed on top of the original exhaustive search
: data-stationarity-based \textit{heuristic search}, data-reuse-based \textit{heuristic search} and early-cost-evaluation-based \textit{iterative search}. 
%These proposed search strategies allow to find the optimal schedule up-to 16$\times$ faster.

Thirdly, an \textbf{Architecture Generator (Section 5)} is built on the top to construct different DNN accelerator architectures, especially focusing on the auto-generation of all valid memory hierarchies under given area budget.
% \lm{[Pouya, let's keep Section 5, otherwise tooooo many things need to change. I don't want to do that... Let's think some highlights for it and add them?]}
%\mv{Next sentence is not relevant at this stage. So removed it for clarity}
%A \textit{memory pool} is required for indication of applicable memory modules, each with certain size, word-length, per-word access energy, and area, providing key information for hardware cost estimation. 

Framework validation (Section 6) and three case studies (Section 7) at different design abstraction levels are conducted to assess the accuracy of the Hardware Cost Estimator, to show the strength of the uneven-mapping supportive Temporal Mapping Generator, as well as to gain insight in the vast design space,utilizing the fully-flexible Architecture Generator.

% ----\input{2_related_work}------
\begin{figure*}[t]
\centering
\includegraphics[width=6in]{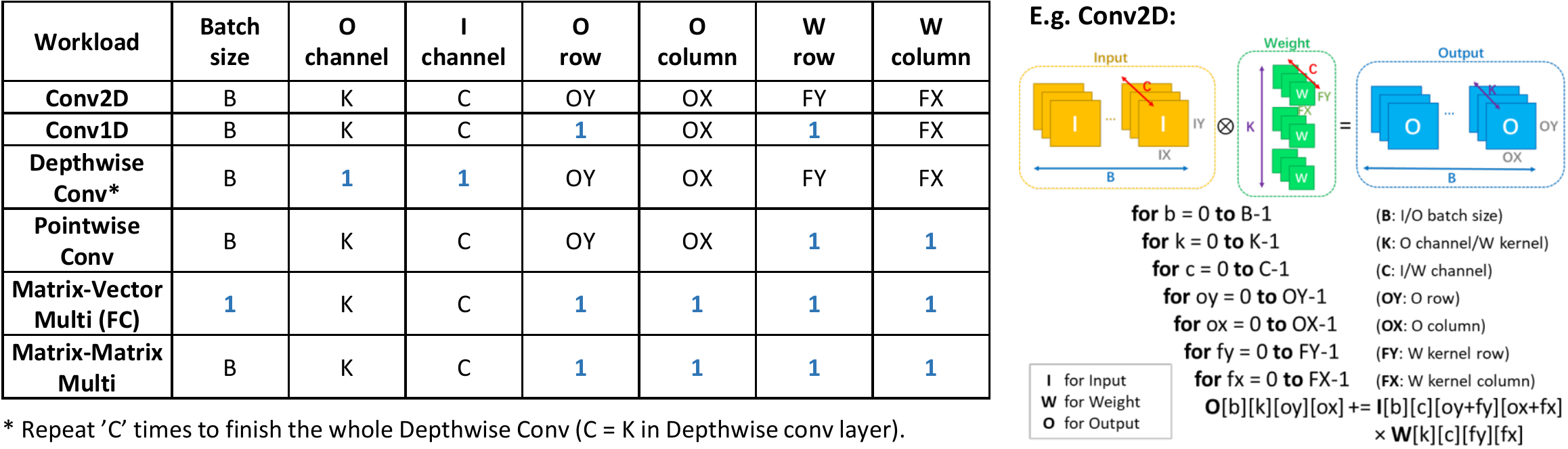}
% \vspace{-1.2em}
\caption{Commonly used neural network workload summary.}
\vspace{-0.8em}
\label{fig:workload}
\end{figure*}

% ----\input{3_overview_and_design_space}------

%\newpage
\section{ZigZag Framework Overview}
\label{section_design_space}

Weight, Input, and Output are the three main operands in each DNN layer. The memory hierarchy and the data mapping scheme (dataflow) are responsible to get Weights and Inputs as efficiently as possible into the multiply-accumulate (MAC) units and collect the resulting Outputs, while maximizing data reuse in local storage. 
%thus reducing memory access from higher \& expensive memory level 
%Towards this goal, a good memory hierarchy and data mapping scheme are highly required. 

However, the many degrees of freedom involved in designing a memory hierarchy and dataflow makes finding the optimal solution a difficult task: 1) Weight, Input, and Output can have the same or different memory organization, 2) for each memory level, Weight, Input, and Output can be stored shared or separately, 3) spatial unrolling can be applied at different memory levels (e.g. register file can be spatially unrolled into each processing element), and 4) for each memory hierarchy, hundreds of thousands of possible schedules exist that strongly impact energy and latency. For these reasons, an automatic tool which can rapidly explore the vast design space becomes a necessity.
%extending the SotA towards this richer memory design space. 

%\mv{This section is very short and I feel it does not add that much new information... That makes it a bit a waste of your valuable space... What you could do is: 1.) shorten the introduction to have more new info in this section 2; 2.) merge this overview section with the SotA discussion, again to lengthen this section, and to discuss the SotA better. Subsection 2.1 could be SotA discussion and subsection 2.2 the framework overview in light of the SotA. Then you will have more balanced section lengths.}
%\mv{Is this framework overview, we ideally also give some additional information regarding the novelty across sections 3-4-5. E.g. the concept of the throughput estimation: where does it impact; the concept of the different memory hierarchies: where does it impacts the framework. Make it even more clear where you differ from SotA. Not in which details (that is already present in figure 1), but especially more on the high level goals. } 

ZigZag targets the automatic exploration of all valid memory hierarchies, given an area constraint, workload size information, and a memory pool. %, which includes all user-defined applicable memory modules. 
This requires an extension of the SotA frameworks on several aspects.

ZigZag contains three key components: 1) an enhanced analytical Hardware Cost Estimator, 2) an efficient and flexible Temporal Mapping Generator, and 3) a memory-centric Architecture Generator. Together, they discover Pareto-optimal design points, each including accelerator architecture, the algorithm's best schedule, and corresponding hardware cost, as shown in Figure~\ref{fig:framework_diagram}.

Besides working in the full-function mode, in which the design space of both architecture and schedule are fully explored, ZigZag can also work in several partial-function modes, in which architecture and/or schedule can be partially or fully pre-defined. For example, if a designer wants to constrain stationarity in the inner-PE to output stationarity to assess the impact of the other factors for this specific architecture, it is possible to freeze degrees of freedom of the DSE and only open the upper levels to the tool to explore. This will be illustrated in several case studies in Section~\ref{section_case_studie}.

% ----\input{4_analytical_model}------

\begin{figure}[h!]
\centering
\includegraphics[width=3.3in]{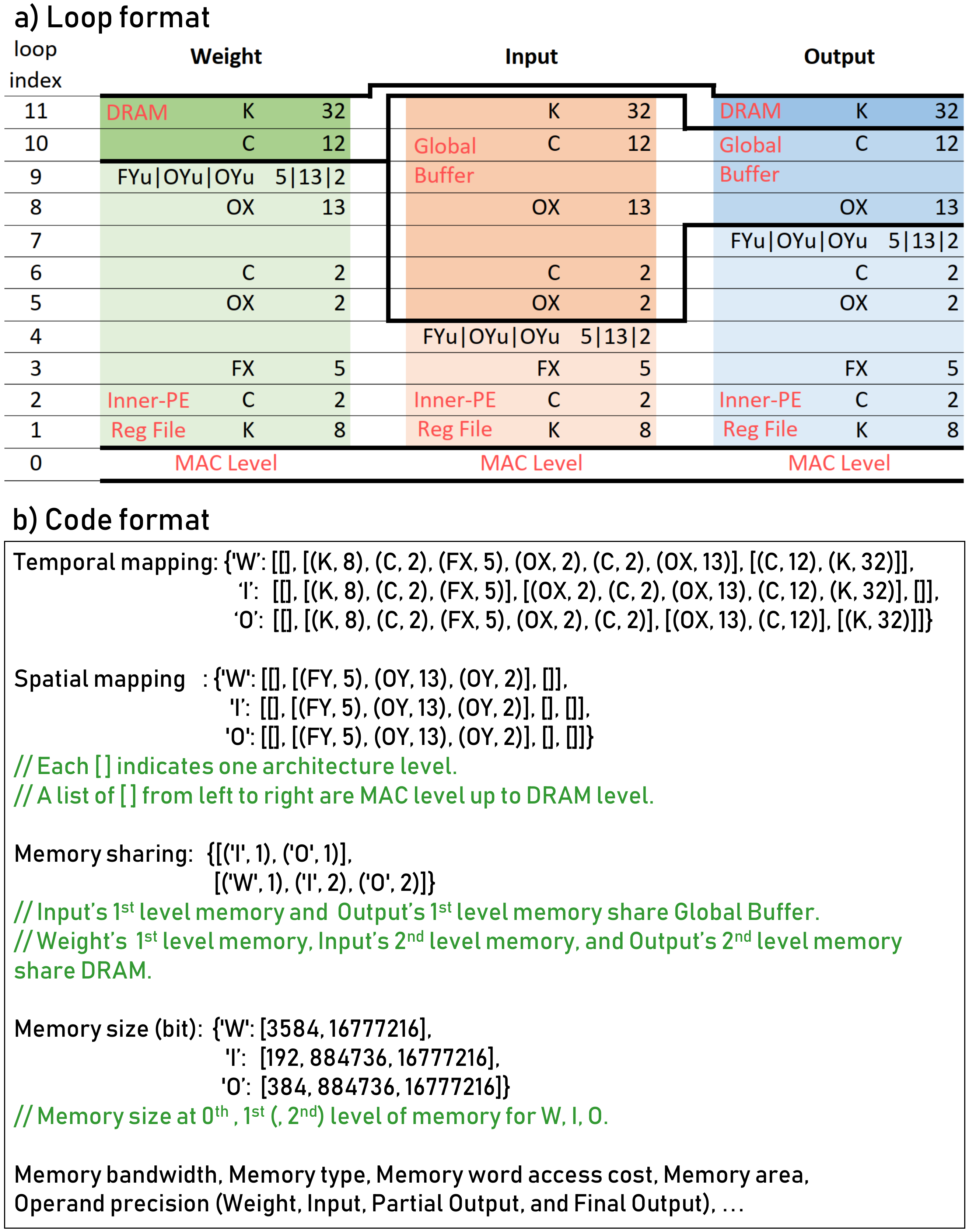}
% \vspace{-1.2em}
\caption{An example of memory-centric dataflow representation (it is the best dataflow out of hundreds of thousands possible dataflows of mapping AlexNet convolutional layer 2 onto Eyeriss V1 architecture~\cite{EyerissV1}). 
%\mv{I do not think it is correct that you use capital letters here for e.g. C and K and ... In figure 2 a captital letter means the loop bound. Here it is the loop itself?  Also, you now have 2 C's and 2 K's. Wouldn't it be better to have k1 and k2, etc? }
}
\vspace{-1em}
\label{fig:loop_demo}
\end{figure}

% \vspace{-2.2em}

\begin{table*}[]
\centering
\caption{Loop relevance's implication on dataflow}
\label{table_loop_impact}
\scalebox{1}{
\begin{tabular}{|c|c|c|c|c|c|c|c|c|c|}
\hline
 & \multicolumn{3}{c|}{\textbf{\textcolor{blue} {\cmark} r loop}} & \multicolumn{3}{c|}{\textbf{\textcolor{red} {\xmark} ir loop}} & \multicolumn{3}{c|}{\textbf{\textcolor{green} {?} pr loop pairs}} \\ \cline{2-10} 
\multirow{-2}{*}{\textbf{\begin{tabular}[c]{@{}c@{}}Loop \\ Impact\end{tabular}}} & \textbf{Spatial} & \textbf{\begin{tabular}[c]{@{}c@{}}Spatio-\\ temporal\end{tabular}} & \textbf{Temporal} & \textbf{Spatial} & \textbf{\begin{tabular}[c]{@{}c@{}}Spatio-\\ temporal\end{tabular}} & \textbf{Temporal} & \textbf{Spatial} & \textbf{\begin{tabular}[c]{@{}c@{}}Spatio-\\ temporal\end{tabular}} & \textbf{Temporal} \\ \hline
\textbf{W} & Unicast & Unicast & \begin{tabular}[c]{@{}c@{}}Fetch\\ new Weight\end{tabular} & Broadcast & \begin{tabular}[c]{@{}c@{}}Propagate\\ systolically\end{tabular} & \begin{tabular}[c]{@{}c@{}}Keep\\ stationary\end{tabular} & \cellcolor[HTML]{C0C0C0} & \cellcolor[HTML]{C0C0C0} & \cellcolor[HTML]{C0C0C0} \\ \hline
\textbf{I} & Unicast & Unicast & \begin{tabular}[c]{@{}c@{}}Fetch \\ new Input\end{tabular} & Broadcast & \begin{tabular}[c]{@{}c@{}}Propagate\\ systolically\end{tabular} & \begin{tabular}[c]{@{}c@{}}Keep\\ stationary\end{tabular} & \begin{tabular}[c]{@{}c@{}}Broadcast\\ diagonally\end{tabular} & \begin{tabular}[c]{@{}c@{}}``FIFO \\ Effect"\end{tabular} & \begin{tabular}[c]{@{}c@{}}``FIFO \\ Effect"\end{tabular} \\ \hline
\textbf{O} & Uni-collect & Uni-collect & \begin{tabular}[c]{@{}c@{}}Generate \\ new Output\end{tabular} & \begin{tabular}[c]{@{}c@{}}Sum up \\ spatially\end{tabular} & \begin{tabular}[c]{@{}c@{}}Accumulate\\ systolically\end{tabular} & \begin{tabular}[c]{@{}c@{}}Accumulate\\ stationarily\end{tabular} & \cellcolor[HTML]{C0C0C0} & \cellcolor[HTML]{C0C0C0} & \cellcolor[HTML]{C0C0C0} \\ \hline
\end{tabular}
}
\end{table*}

\section{Hardware Cost Estimator}
\label{section_ce}
The ZigZag Hardware Cost Estimator targets the estimation of energy and performance (PE array utilization, throughput, and latency) given a certain workload size, dataflow (temporal and spatial mappings), and memory hierarchy.

It innovates on 1) Memory-centric dataflow representation (Section~\ref{sub_section_represent}) to capture {the interaction between} dataflow and memory hierarchy; 2) a loop relevance principle (Section~\ref{sub_section_relevance}), to extract basic technology-independent hardware and data attributes from loop sets, such as memory access count and required memory bandwidth; 3) a technology- and memory-bandwidth-aware hardware cost integrator (Section~\ref{sub_section_cost}), capable of not only extracting energy, but also latency. 
%\mv{I rewrote a bit, trying to highlight innovation and differentiators a bit more.}

\subsection{Memory-Centric Dataflow Representation}
\label{sub_section_represent}
%\mv{The next two paragraphs are not specific for the cost estimation. It is more general, and hence does not belong to section 3. It belongs to section 2, or even 1?}

A uniform and concise data representation format lays the foundation for the exploration of the enlarged design space and is required to support all forms of memory sharing, loop blocking (loop tiling), loop ordering, and spatial unrolling for each operand (W/I/O) at each memory level. 
The proposed \textit{Memory-Centric Dataflow Representation} well captures all memory hierarchy attributes as well as spatial and temporal algorithm-to-hardware mapping schemes.

Figure~\ref{fig:loop_demo} illustrates the proposed memory-centric dataflow representation using the same loop name notation as Figure~\ref{fig:workload}.
%\mv{figure 2 should not be introduced here, but earlier...}.
The depicted dataflow is the energy-optimal dataflow found by ZigZag out of hundreds of thousands of possible dataflows for mapping AlexNet convolutional layer 2 (with B=1) onto Eyeriss V1 architecture~\cite{EyerissV1}\footnote{In this paper, We adopt the AlexNet layer dimensions reported in Eyriss paper~\cite{EyerissV1} for fair comparisons, which are different to the layer dimensions reported in AlexNet paper~\cite{AlexNet}.}, leading to 20\% energy savings compared to the original dataflow used. 
Notice that in this example, the memory hierarchy and spatial unrolling settings are the same as Eyeriss, while only the temporal mapping is different, i.e. different loop blocking and loop ordering.
%\lm{We cannot remove it because it is super important information for this example.}
%\mv{I removed some info on Eyeriss here. It distracts (just like the first 2 paragraphs of this sub-section) from the explanation you actually have to do here, which is your representation. Alternative could be to introduce this example already in section 2 or in the intro of this section 3.}
This example will be used throughout this paper to explain various aspects of the framework. 

In Figure~\ref{fig:loop_demo}a, the representation defines, from left to right, the dataflow information of the three operands separately, using three sets of nested for-loops. Inside each set, the architectural levels are represented from bottom to top (divided by bold lines), starting from the MAC units, over potential register file and/or SRAM (Global Buffer) levels, all the way up to DRAM. 
%\mv{Note: You have to explain here what the bold lines mean. You can at the same time explain what loop blocking is (will you need later, see my later comment.)\\}
For each operand, each alphanumeric pair indicates a for-loop, e.g. the first term ``\texttt{K 32}" is equivalent to ``\texttt{for k = 0 to 32-1}". Assigning these for-loops into different architectural level is loop blocking (loop tiling) and fixing the order of all the for-loops inside one level is loop reordering. The ``\texttt{u}" suffix after a loop name indicates spatial unrolling, such as ``\texttt{FYu}". The format ``\texttt{Au|Bu}" is inherited form \cite{Yang2018}, meaning that both the A and B loop dimensions are spatially unrolled. 
%\mv{should we also not add that  all temporal unrollings must be identical across all 3 operands, but spatial can be at different locations?  Or is this an observation that should be added to the 3 observations below?  Use the sentences from you below I indicated there that they can be moved up.} 
%\mv{You have to explain figure 3b first, before you can use it in the reasoning below. Are 3a and 3b equivalent? Or 3b has more info? Does the tool only uses 3b, or also 3a?  If you can explain every with only 3a, then give the key reasonings first, and only afterwards explain what 3b is and why you need it...}
In Figure~\ref{fig:loop_demo}b, more detailed information is given, in which temporal mapping (schedule), spatial mapping, memory sharing, memory size, etc. are specified for each operand at each architectural level.

By combining Figure~\ref{fig:loop_demo}a with \ref{fig:loop_demo}b, three key attributes of this representation can be observed:
1) not all of the operands have the same number of memory levels, e.g. in this example Weight has two memory levels while Input and Output have three; 2) not all of the operands that have the same memory level share physical memory, e.g. the Inner-PE Register File of W/I/O are separated; 3) not all of the operands that share a physical memory have the same/even loop blocking, e.g. Input and Output share the Global Buffer, but with a different loop blocking boundary.

Furthermore, it can be noted that temporal loops of all operands should follow the same order to maintain functional equivalence, while spatial loops can be relocated. This is because spatial mapping in this representation indicates which loop dimension is unrolled at which architecture level and to what extent, which is fully configurable in ZigZag. In this particular example, following the Eyeriss settings, W, I, and O have the same spatial mapping at the same memory level, i.e. ``\texttt{FYu|OYu|OYu 5|13|2}" at Register File level.
%\mv{Huh? This confuses me. I see these loops at different levels, e.g. nb 9 for W and nb 4 for input?!}
%\lm{Then are at the same memory level.}

\subsection{Loop Information Extractor Based On the Loop Relevance Principle}
\label{sub_section_relevance}

\begin{figure}[t]
\centering
\includegraphics[width=3.3in]{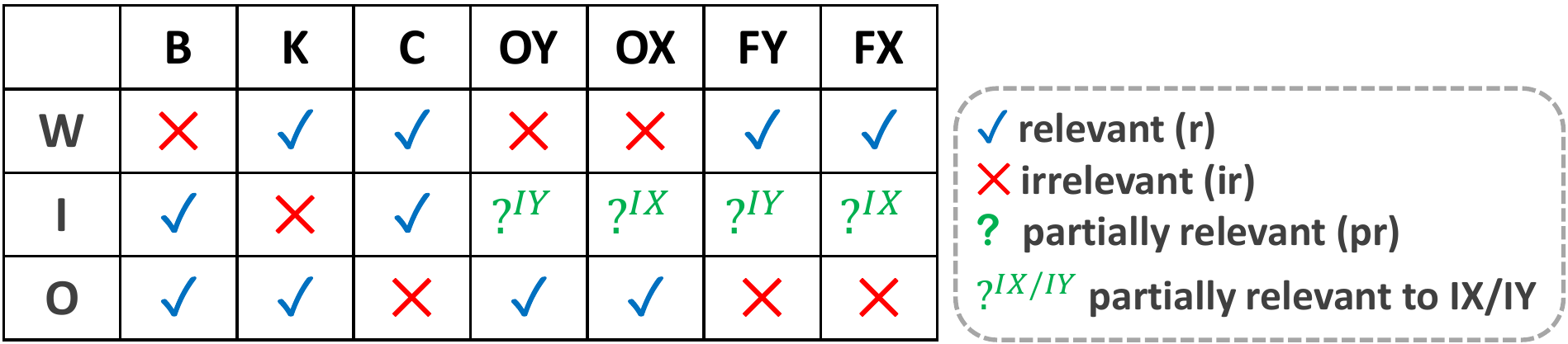}
% \vspace{-1.4em}
\caption{Loop type categorized by relevance.}
% \vspace{-1.4em}
\label{fig:loop_type}
%\vspace{-1.6em}
\end{figure}

The enhanced representation from Section 3.1 will now be combined with a loop relevance principle, to systematically analyze and extract basic technology-independent hardware and data attributes.

Convolutional layers are based on a 7D computing space with three 4D operands:  Weight, Input, and Output; which implies not all 7 dimensions are relevant to each operand.
%Convolutional layer intrinsically has a 7D computing space with three 4D operands, Weight, Input, and Output, thus not all 7 dimensions are relevant to each operand. 
Figure~\ref{fig:loop_type} shows the loop relevance principle, in which all 7 loop dimensions are categorized as relevant (r), irrelevant (ir), or partially relevant (pr) to each operand. For Weight and Output,
%\seb{comment: the prhase "things are straightfoward" can be replaced by a more formal way} \ns{Dataflow for Weight and Output are straightforward since..} 
this is straightforward since all 7 computing space dimensions are either parallel (relevant) or orthogonal (irrelevant) to their own 4D data space. Looping through those `r' loops indicates new data need to be fetched or generated, while looping through those `ir' loops creates various data reuse opportunities, as shown in Table~\ref{table_loop_impact}.

%\vj{This whole para needs some work, try to make it easy to understand}
%\seb{not clear phrase} 
%\seb{comment: Maybe you can put the key words like "Input" in italic to differentiate it}
Input, however, also has `pr' loops besides the `r' and `ir' loops. As presented in the right example of Figure~\ref{fig:workload}, Input's dimensions \texttt{IX} and \texttt{IY} do not show up in the convolution formula directly, instead they are indirectly present through \texttt{OX} and \texttt{FX} (for \texttt{IX}); \texttt{OY} and \texttt{FY} (for \texttt{IY}). As such, \texttt{OX}, \texttt{FX}, \texttt{OY}, \texttt{FY} are denoted as partially relevant (pr) loops for Input. \texttt{OX}, \texttt{FX} (resp. \texttt{OY} and \texttt{FY}) form a `pr' loop pair. For a `pr' loop pair, data reuse opportunities arise when the sum of their indices remains constant while the computation is looping through its space.

\begin{figure}[t]
\centering
\includegraphics[width=3.5in]{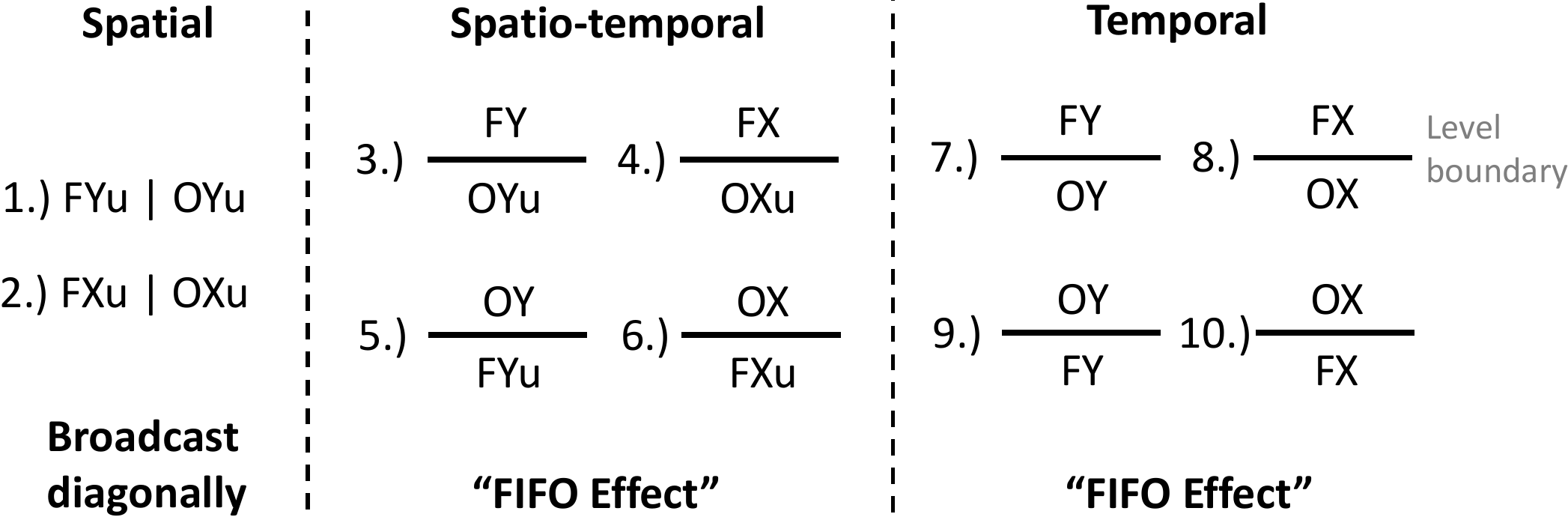}
\caption{Partial-relevant (pr) loop patterns that trigger special Input data reuse.}
\label{fig:FIFO_effect}
\vspace{-1em}
\end{figure}

Figure~\ref{fig:FIFO_effect} provides a summary of pr-loop-pair-triggered input data reuse. 
Such `pr' creates alternative data reuse opportunities for spatially, temporally or spatio-temporally unrolled loops. For spatial unrolling, inputs can be broadcasted diagonally in a PE array, as done in Eyeriss~\cite{EyerissV1}, where \texttt{FY} and \texttt{OY} are spatially unrolled onto the 2D PE array, allowing a diagonal broadcast of inputs. For temporal and spatio-temporal unrollings, 
%\mv{\textit{Note: I have the feeling we did not well explain what we mean with spatio-temporal?}} 
data reuse is possible through a FIFO buffer which shifts the input data over consecutive clock cycles. An example of this can be found in Envision~\cite{ENVISION}, where \texttt{OX} is spatially unrolled and \texttt{FX} is the innermost temporal loop on top, same as Figure~\ref{fig:FIFO_effect} (4), 
making the sum of \texttt{FX} and \texttt{OX} a constant in neighboring PE locations across consecutive cycles, enabling to reuse Inputs in a FIFO manner.%, which we named Input ``FIFO Effect". 

\begin{figure*}[]
\centering
\includegraphics[width=\linewidth]{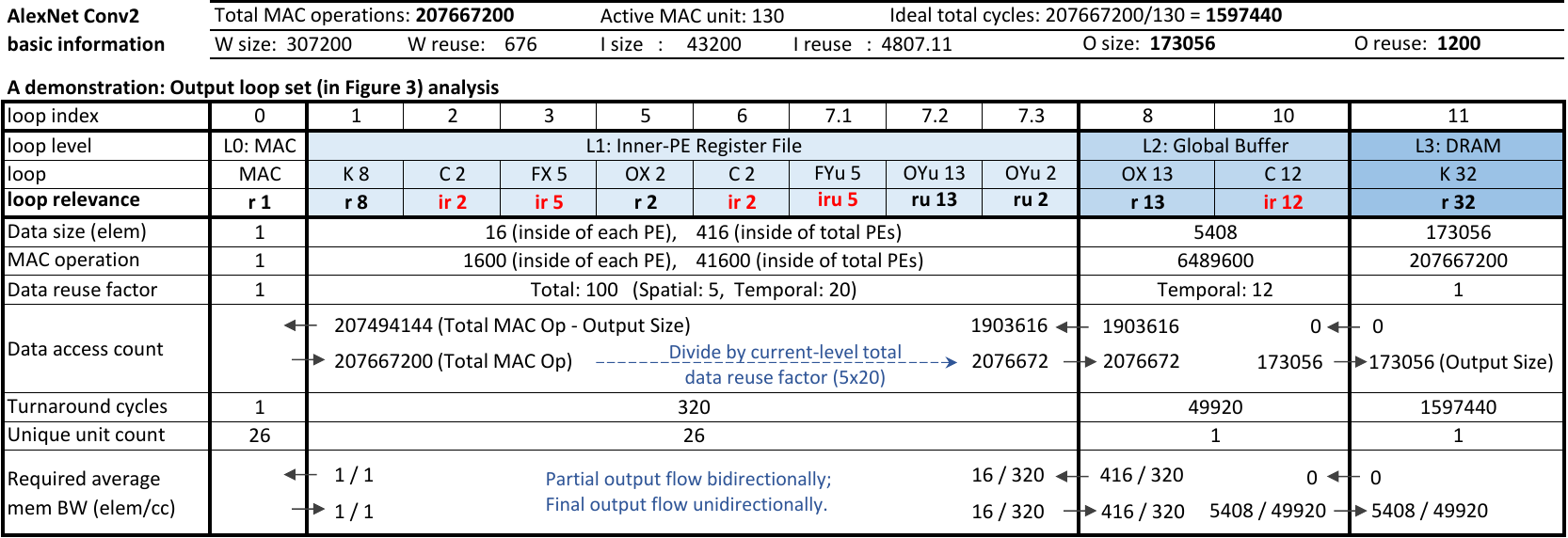}
% \vspace{-1.2em}
\caption{A demonstration: extract loop information from Output loop set in Figure~\ref{fig:loop_demo} based on loop relevance principle.}
% \vspace{-1.0em}
\label{fig:extractor_demo}
\end{figure*}

The benefit of this loop relevance principle is the simplification and unification of the procedure for extracting key information from the W/I/O loops sets towards estimating system energy and performance. To show the key ideas of this procedure, an equation summary is provided in Table~\ref{fig:Equation} and a detailed demonstration is given in Figure~\ref{fig:extractor_demo}, in which the Output loop set is analyzed (the same/similar procedure would be repeated for Weight/Input, not shown).
%mv{I feel you miss explaining that these computations are done once per memory level, and what you mean by a memory level?}

\textit{1.) Data Size in individual memory unit at current level} can be derived by multiplying together the dimensionality of all the `r' loops at the current level and all levels below, together with all `ru' loops (spatially unrolled `r' loop) at all levels below. This can be seen in the first line of Table~\ref{fig:Equation}, in which $Li$ means current memory level, $L(i-1)$ means one level below the current memory level, and $Lmin$ means the lowest memory level. 
Let us apply this to a specific example, given in Figure~\ref{fig:extractor_demo}. The required Output data storage inside each PE (16) is calculated by multiplying the dimensionalty of Level-1 `r' loops 1 and 5; the Data Size of the Output inside of Global Buffer (5408) is calculated by multiplying the dimensionality of Level-1,2 `r' loops (1, 5, 8) and Level-1 `ru' loops (7.2, 7.3). Later for other metrics calculation, readers can always refer to the practical case in Figure~\ref{fig:extractor_demo} for validation.

\textit{2.) Data Size in total at current level} can be easily calculated by multiplying the individual Data Size in each memory unit with the dimensionality of all `ru' loops at current level. 
Notice that the unit of the Data Size is number of elements. In order to obtain number of bits, the precision of the operands needs to be considered. Generally speaking, partial outputs have a higher precision than weights, inputs, and final outputs. 
%\seb{flowing bidirectionally between XXX and XXXX} 
%and it flows bidirectionally. 
%\lm{I have explained it in the figure}
The ability to distinguish partial outputs
%, which need writes and reads, 
from final outputs
%, which need only writes, 
is critical in the framework for accurate hardware cost estimation. ZigZag can easily handle through its `r' vs `ir' loop representation. The final output is generated at the level of the uppermost 'ir' loop., e.g. the L2 Global Buffer in Figure~\ref{fig:extractor_demo}. As such, the data traffic between L2 and L3 is unidirectional.
%(write only).

\begin{table}[t]
\centering
\caption{Equations for loop information extraction} %\mv{to save space, maybe remove the less important equations that you do not explain, like e.g. "data size in total" and "turnaround cycles", and "unit count". In that way, you save space in the table, as well as in the text, and you create less confusion.}}
\includegraphics[width=3.3in]{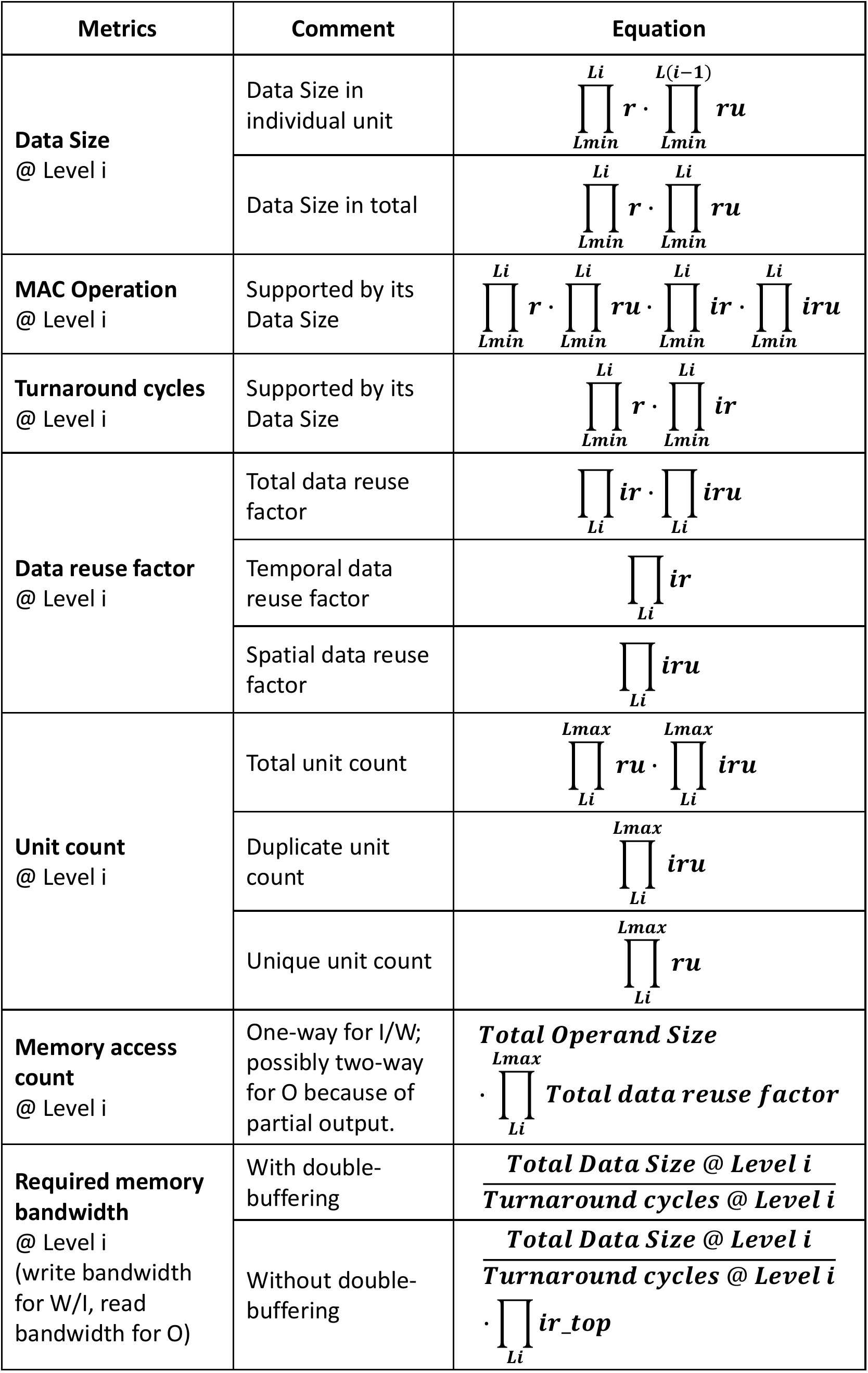}
\label{fig:Equation}
% \vspace{-2.9em}
\end{table}

\textit{3.) Number of MAC Operation} supported by current level Data Size is calculated by multiplying together all the loops' dimensionalty (`r', `ir', `ru', and `iru') from the lowest level up to the current level.

\textit{4.) Turnaround cycles} are number of cycles certain memory can keep operating with the data it contains, which is an important metrics for later required memory bandwidth computation. It can be calculated by multiplying together all the temporal loops' dimensionalty (`r' and `ir') from the lowest level up to the current level.

\begin{figure}[t]
\centering
\includegraphics[width=3.3in]{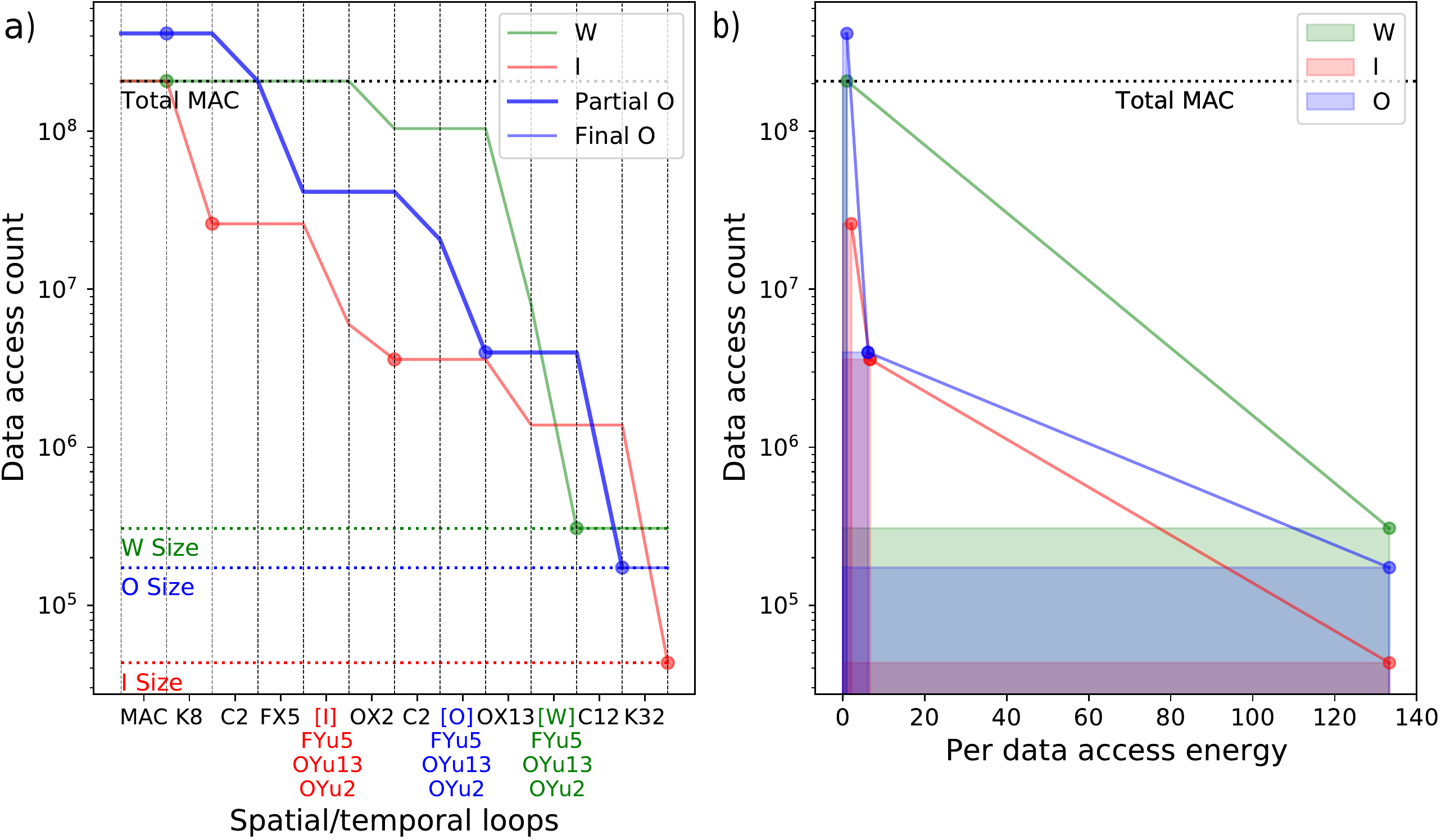}
% \vspace{-1.5em}
\caption{Visualization of a) the impact of individual loops (in a bottom-up order) on data access count, and b) energy consumed by different memory levels (the area of these blocks indicates energy), using the dataflow example in Figure~\ref{fig:loop_demo}. Note that the highlighted dots in two figures are one-to-one correspondence and are corresponding to the data access at different memory levels for different operands.}
% \vspace{-1.5em}
\label{fig:loop_vis}
\end{figure}

\textit{5.) Total data reuse factor} at current level is the product of all the irrelevant loops' dimensionalty (`ir' and `iru') at current level. The product of only `ir' loops is the temporal data reuse factor, while the product of only `iru' loops is the spatial data reuse factor.

\textit{6.) Total unit count} is a metrics that measures how many hardware components are at certain level, which is only related to spatial unrolled loops. Total unit count at current level is the product of all the spatial loops' dimensionalty (`ru' and `iru') from the current level up to the highest level.

\textit{7.) Duplicate unit count} is a measurement of how many hardware components at certain architectural level that contain same data, which is captured by the product of all the irrelevant spatial loops' dimensionalty (`iru') from the current level up to the highest level. 

\textit{8.) Unique unit count} is similar to Duplicate unit count, but for counting how many hardware components that contain different data, thus all the relevant spatial loops' dimensionalty (`ru') are timed together from the current to highest level.
%It need to be taken into account for computing the above-level required memory bandwidth.

\textit{9.) Memory access count}, as the core metrics for later memory energy estimation, can also be easily extracted. The first term in the formula, ``Operand Size" is how many elements in total W, I, or O has; the second term is how many times each element needs to be accessed repetitively at the current memory level, which equals to the product of the total data reuse factor at current level and all the levels above. Figure~\ref{fig:loop_vis}a visualizes individual loop's impact on the memory access count. The circle markers indicate the boundary of the memory levels, showing the actual number of memory accesses for each memory level for each operand.

\textit{10.) Required memory bandwidth} is the minimum bandwidth that ensures computation happen fluently without stall. It depends on both dataflow and memory settings. Without double-buffering, writing only happens after a specific data item is fully used, resulting in a small time window. With double buffering, writing can happen all the time (in parallel with data loading), leaving a large writing time window, and thus lowering required instantaneous memory bandwidth. The bandwidth difference between these two cases is the product of all the top `ir" loop values.

Note that due to the `pr' loops, some changes are needed for handling the Input. The most important modification are the following two substitutions. One is to correctly handle data size (assuming stride is 1):
\begin{small}
\begin{equation}
\notag
\prod_{Lmin}^{Li}{r}\rightarrow\prod_{Lmin}^{Li}{r}\cdot(\prod_{Lmin}^{Li}{pr_1}+\prod_{Lmin}^{Li}{pr_1'}-1)\cdot(\prod_{Lmin}^{Li}{pr_2}+\prod_{Lmin}^{Li}{pr_2'}-1)
\end{equation} 
\end{small}
\\in which ${pr_1}$ (${pr_2}$) and ${pr_1'}$ (${pr_2'}$) are a pr loop pair, like \texttt{OX} and \texttt{FX}. Another substitution is to correctly handle special Input data reuse cases like the ``diagonal broadcast" and ``FIFO Effect": 
% \vspace{-0.5em}
\begin{equation}
\notag
{\text{Total\;data\;reuse\;factor\;@\:Li}}\rightarrow\frac{\text{Total\;MAC\;Op\:@\:Li(+pr)}}{\text{Total\;Data\;Size\:@\:Li(+pr)}}
\end{equation} 

%In which, to handle special Input data reuse cases correctly like ``diagonal broadcast" and ``FIFO Effect", we use the total number of MAC operations divided by the total input data size, at that level PLUS the above-level special-data-reuse-triggering pr loop (if it exists). 
For example, in the ``FIFO Effect" setting: $\frac{\text{FX\;3}}{\text{OXu\;4}}$, the lower-level data reuse factor should equal to $\frac{(3\times4)\;\text{MAC\;Op}}{(3+4-1)\;\text{data}} = 2$ instead of $\frac{(1\times4)\;\text{MAC\;Op}}{(1+4-1)\;\text{data}} = 1$ by taking the ``FIFO effect"-triggering `pr' loop ${\text{FX\;3}}$ into account.

\subsection{Hardware Cost Integrator}
\label{sub_section_cost}
The Hardware Cost Integrator aims at integrating the extracted technology-independent loop information with the technology-dependent characteristics to estimate the final hardware cost and performance, namely energy, throughput, and area.

\textit{1.) Area:} Area estimation is straightforward, summing up all the used on-chip memory's area, as it is dominant.

\textit{2.) Energy:} MAC computation energy and memory access energy are taken into account. MAC computation energy is estimated by multiplying total number of MAC operations with average single-MAC-operation energy; memory access energy is calculated by multiplying the memory access count, provided by the Loop Information Extractor, with the corresponding memory per-data-access energy, taking into account the memory size, the potential memory bitwidth mismatch overhead, operand precision, and data stationarity. Figure~\ref{fig:loop_vis}b visualizes this step: the energy consumed by each memory level  (L1 register file, L2 global buffer and L3 DRAM) of each operand is visualized by the area of each block, showing that a good dataflow leads to high data access counts at low-cost memories with low data access counts at high-cost memory. A reliable wire cost model for interconnection energy estimation is planned for future work.

\textit{3.) Latency/Throughput:} 
%For performance estimation, ZigZag is able to estimate latency for certain workloads.
%, taking workload size, dataflow, memory type, and memory bandwidth into account. 
PE array utilization, throughput, and latency are tightly related and can be deduced from each other. A PE array's under-utilization can come from spatial stalls and temporal stalls. Spatial stalls result from mismatch between the spatial unrolling dimensions and the neural network layer dimensions. Temporal stalls mainly come from memory bandwidth bottlenecks during computation.

ZigZag analytically estimates both types of stalls. Spatial stalls are straightforward. Temporal stalls are calculated by comparing the actual memory bandwidth with the required memory bandwidth (derived from Loop Information Extractor), which is tightly coupled to memory type and dataflow. 

\begin{figure*}[t]
\centering
\includegraphics[width=6.8in]{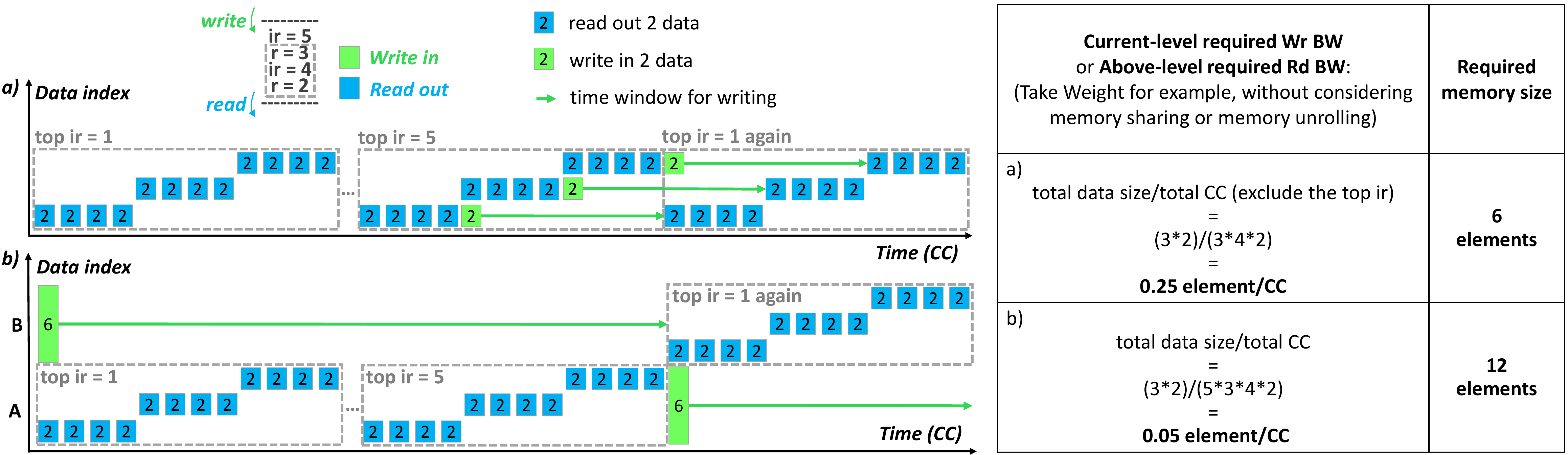}
\caption{ir-loop-on-top dataflow's implication on required memory bandwidth and memory size with a) non-double-buffering and b) double-buffering. Assuming a) is with one r/w dual-port memory and b) is with two single-port memories. ``CC" stands for clock cycle; ``BW" stands for bandwidth.}
% \vspace{-0.5em}
\label{fig:loop_latency_db}
\end{figure*}

Figure~\ref{fig:loop_latency_db} gives an toy example of extracting required memory bandwidth with two memory scenarios under a memory level with a `ir' loop on top. In Figure~\ref{fig:loop_latency_db}a), without double-buffering, writing only happens when one datum is fully used, thus the time window left for data writing is small, while in Figure~\ref{fig:loop_latency_db}b), with A and B buffer, writing can be totally overlapped with reading, leaving a large time window.
%, thus a lower memory writing bandwidth is required. 
More specifically, the required writing memory bandwidth for case a) is $6/24=0.25$, for case b) is $6/120=0.05$. The ratio between them is exactly the size of top-ir loop.

\begin{figure}[h]
\centering
\includegraphics[width=3.5in]{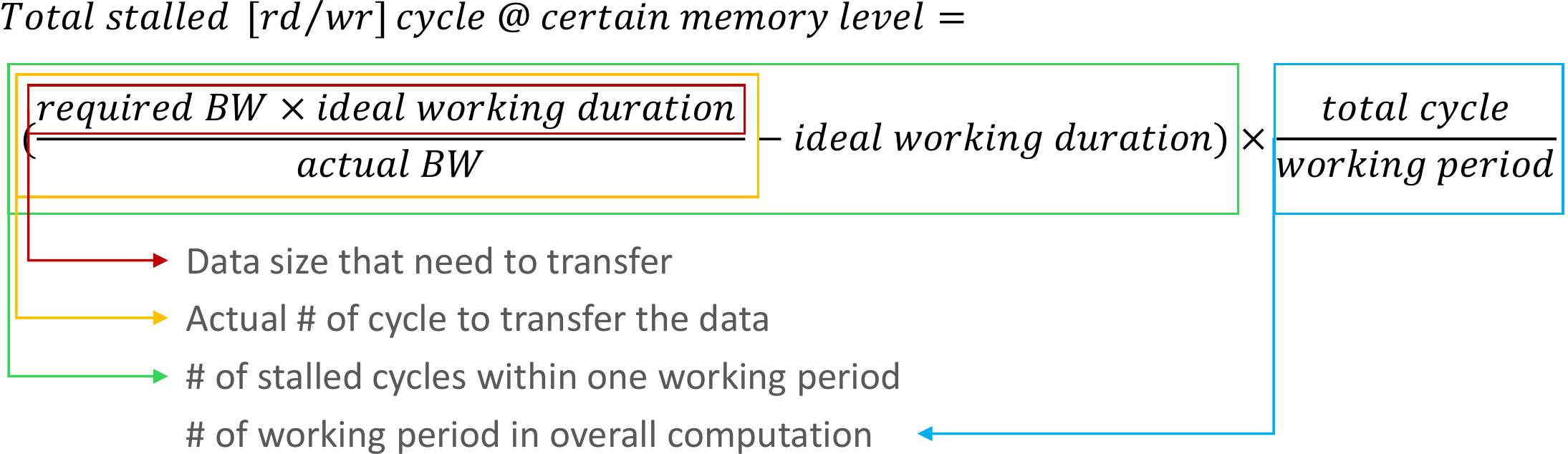}
\caption{Stalls due to limited memory bandwidth.}
% \vspace{-0.8em}
\label{fig:latency}
\end{figure}

After getting the required memory bandwidth, the next step is analyzing ideal memory data transfer duration and data transfer period, i.e. understanding how long and how often one memory is working. Then, stalls due to limited memory bandwidth can be calculated by the equation shown and explained in Figure~\ref{fig:latency}.

% \begin{figure}[h!]
% \centering
% \includegraphics[width=3.3in]{figures/latency3.pdf}
% % \vspace{-1.2em}
% %\caption{Formula for stalling cycles caused by memory bandwidth limitation.}
% %\mv{Do not make this a figure (takes too much space), but just a formula in the text which you briefly (not too elaborately) explain.}}
% % \vspace{-0.2em}
% \label{fig:latency}
% \end{figure}

The practical situations are usually more complicated. ZigZag's performance analysis incorporates the following factors: memory bandwidth, memory sharing between W/I/O, memory type (single-port/dual-port, wi/wo double-buffering), memory spatial unrolling, and partial/final sum of output.

\begin{figure*}[t]
\centering
\includegraphics[width=6in]{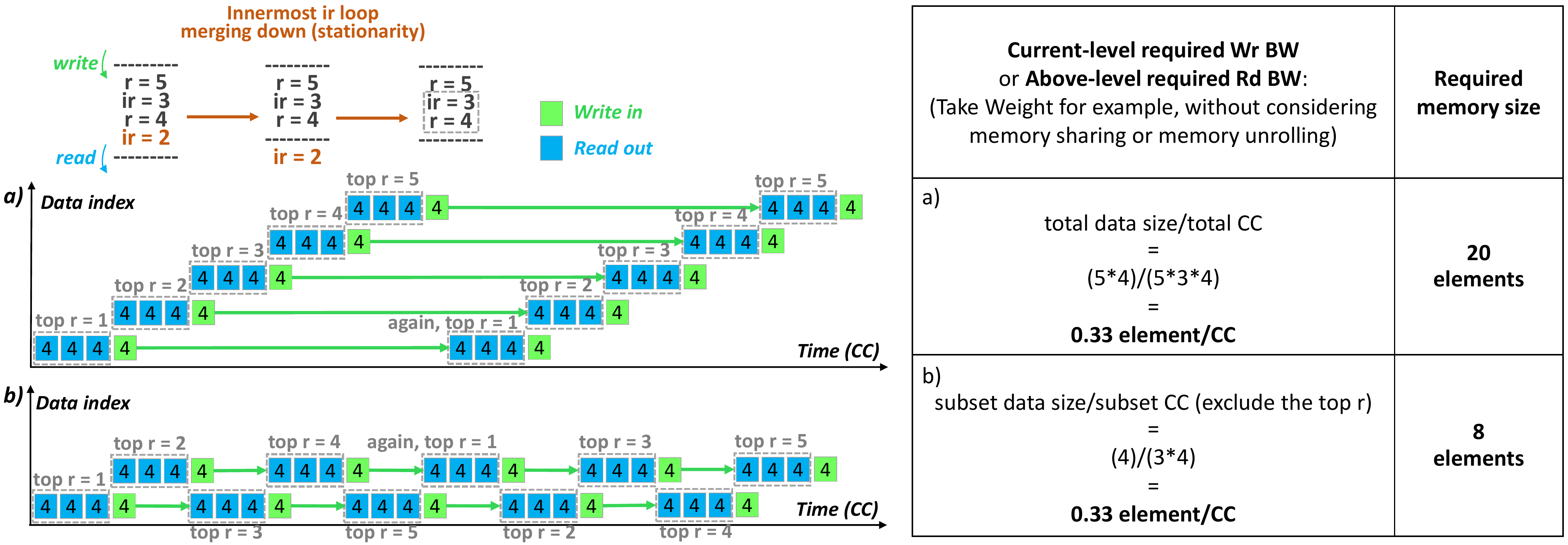}
\caption{r-loop-on-top dataflow's implication on required memory bandwidth and memory size. Assuming both a) and b) are with one r/w dual-port memory.}
% \vspace{-0.5em}
\label{fig:loop_latency0}
\end{figure*}

Besides estimating latency, another potential of ZigZag's performance analysis is detecting run-time memory gating possibilities. For example, Figure \ref{fig:loop_latency0}a) and b) show two valid memory schedules (with the same required memory bandwidth) to support the smooth computation of the same dataflow schedule (written in `r'/`ir' loop format). Notice that b) need less memory size than a), which indicates that, theoretically, 60\% of the memory size in scheme a) can be gated. We call the minimal required memory size to support certain dataflow schedule performing smoothly without affecting the memory bandwidth requirement as ``effective memory size". The memory part that exceeds the ``effective memory size" can theoretically be gated to save power without affecting performance. ZigZag provides the effective memory size analysis for each valid mapping point.

%\mv{Do not make this a figure (takes too much space), but just a formula in the text which you briefly (not too elaborately) explain.}

% ----\input{51_temporal_generator}------
\begin{figure*}[h]
\centering
\includegraphics[width=7.1in]{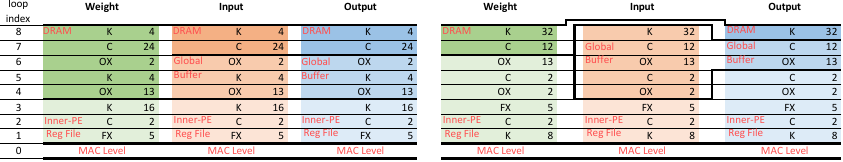}
% \vspace{-1.2em}
\caption{The best \hl{even} temporal mapping of AlexNet CONV2 on Eyeriss (left) compared to the best mapping globally, which is \hl{uneven} (right). The \hl{uneven} one can achieve 20\% energy saving with the same spatial mapping, "FYu$|$OYu$|$OYu  5$|$13$|$2" at Inner-PE Register File level (not shown in the figure).}
% \vspace{-1.2em}
\label{fig:loop_balanced}
\end{figure*}
\section{Temporal Mapping Generator}
\label{section_tmg}
The workload is expressed as a set of nested-loops that determine the order of execution of the MAC operations within each PE. Since the operands for the loops are distributed in the memory hierarchy, each loop is mapped on a level at an index order. The order, size and type of the nested-loops determines the \textit{temporal mapping}.

The ZigZag mapper efficiently searches for loop schemes on \hl{even} and \hl{uneven} memory hierarchies that present shared and/or non-shared levels between different operands. Thus, it \textit{significantly} increases the design space with respect to previous works, without missing the optimal solution. 

%Depending on the complexity of the layer in a given network and on the characteristics of the hierarchy, the number of possible mappings can be unfeasibly large for a single exploration run. %This is due to the fact that the number, the type, the size and the order of nested loops of the optimal mapping can be one of a considerably large amount of combinations. 
By adopting an enhanced loop blocking space representation and using the concept of \textit{roofs} and \textit{virtual levels}, the proposed ZigZag mapper can efficiently support:
\begin{itemize}
\itemsep0em 
\item 2D and 3D convolutional layers, pointwise convolutional layers and fully-connected layers. Depthwise layers can be expressed as a combination of Conv2D and pointwise layers as described in Figure \ref{fig:workload};
\item Memory hierarchies with memory levels that store separate operands and/or memory levels that  are shared with two or three types of operands, so as to allow maximum flexibility in the hardware design space;
\item Multiple levels of spatial unrolling;
\item Three mapping space exploration methods: exhaustive search, heuristically-optimized search and non-heuristically iterative optimized search.
\end{itemize}

\subsection{Enhanced Loop Blocking Representation}

\subsubsection{Loop prime factors}
The module generates all the valid temporal mappings, that are characterized by their type, size and order of the nested loops. In order to fully explore all possible combinations of schemes, we will refer to \textit{loop prime factors} (LPF) as the smallest sizes in which a loop can be split, or in other words, the atomic blocking sizes that cannot be further divided. 

These LPFs correspond to the result of the factorization of the layer dimensions and are the basic blocks of the search algorithm. Starting from the smallest memory level in the hierarchy, the search method proceeds in successive allocations of these prime factors and generates all valid assignment combinations of the LPFs in the given memory hierarchy.

\begin{figure*}[t]
\centering
\includegraphics[width=5in]{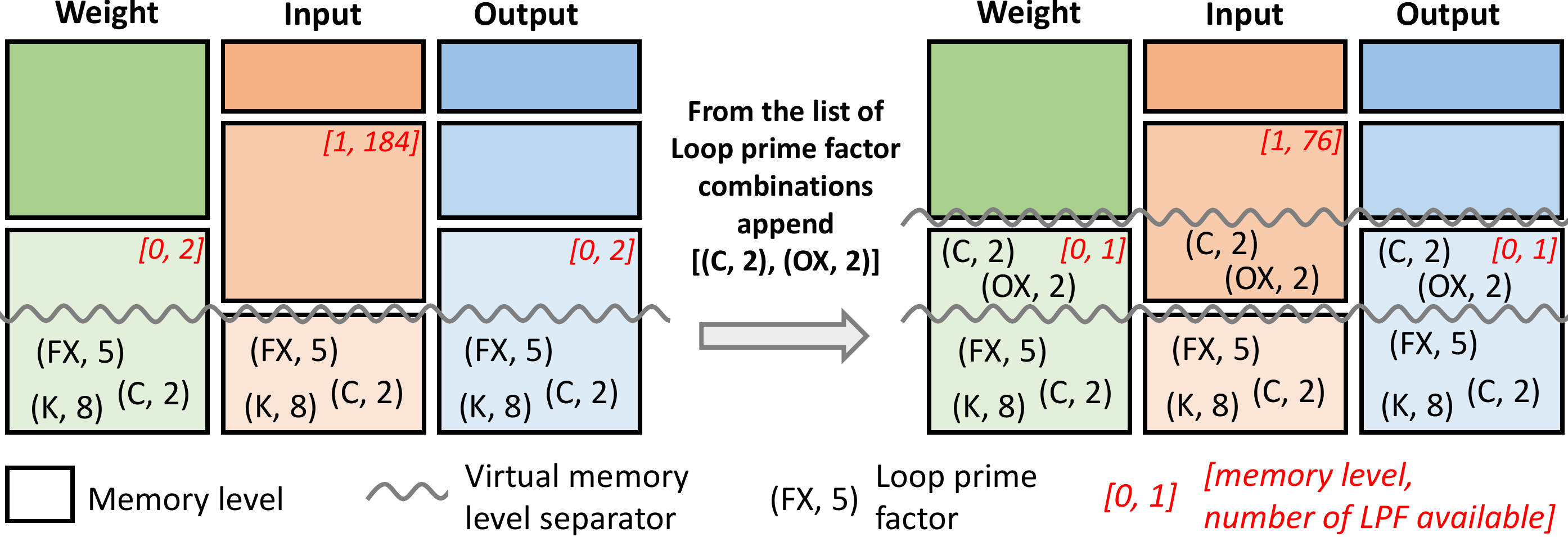}
% \vspace{-1.2em}
\caption{The LPF assignment step in a partial scheme and the successive update of the roof variable. The roof values are those in red: they are defined for each operand and updated after each LPF allocation.}
% \vspace{-1.2em}
\label{fig:loop_blocking}
\end{figure*}

\subsubsection{Roofs and virtual memory levels}
Much of the previously published temporal mapping search methods only dealt with \hl{even} mappings. In those search methods, every for-loop belonged to the same memory level for inputs, weights and outputs, meaning a considerable restriction of the mapping space.
With \hl{even} hierarchies the mapping process of the blockings purely implies a fast check whether the size of the blocking combination and loop order fits within the available space in the corresponding memory level. In contrast, under \hl{uneven} hierarchies, a single blocking index can belong to different memory levels for different operands as in Figure \ref{fig:loop_balanced}. To handle these scenarios, ZigZag introduces \textit{virtual memory levels} and the \textit{roof} variable. %To overcome this issue the unbalanced hierarchy is expressed as a set of \textit{virtual memory levels}, within which equal sets of LPFs are present for all operands.

The assignment process is guided by the \textit{roof} variable, which is a tuple defined for each operand, containing 1) the memory level index where the LPFs are being assigned and 2) the maximum amount of relevant prime factors that can still be assigned in the available space in the specified memory level. The roof is initialized with the smallest memory level in the hierarchy for each operand since the assignment begins from the innermost level of the hierarchy. Its value is updated after each assignment by dividing the available space by the product of the size of the relevant loops allocated. 

A single assignment step is described in Figure \ref{fig:loop_blocking}: the memory hierarchy is the same as Eyeriss and the workload is CONV2 of AlexNet. The figure contains a partial scheme with some LPFs already assigned in the innermost memory levels. The reported partial scheme is one of many other possible ones that have been obtained with previous LPF assignment steps. In Figure \ref{fig:loop_blocking}, before the assignment of the LPFs, the second level of input memory hierarchy (at index 1) has a storage capacity of 884736 bits, as described in Figure \ref{fig:loop_demo} or 884736/16 = 55296 blocks that can be stored with precision 16 bit. The prime factors that this level holds are those already assigned and relevant in the levels below (\texttt{(FX,5)},\texttt{(C,2)}) plus those relative to the relevant spatial unrolling below (\texttt{(FYu,5)},\texttt{(OYu,13)},\texttt{(OYu,2)}) that combined correspond to $5 \times 2 \times (5 + 26 - 1) = 300$ blocks. Therefore, the second level can still store $55296/300 = 184.32 \sim 184$ blocks. The roof for the input in this scheme is thus \texttt{[1, 184]}. 

In the described case, the LPF combination of \texttt{(C,2)} and \texttt{(OX,2)} is one of the many combinations found to be fitting within the roof limits and is appended to the partial scheme. 
After its allocation the available space at the same level becomes $55296 / [(5+2-1)\times 2 \times 2 \times (5 + 26 - 1)] = 76$ blocks and thus the roof will be updated to \texttt{[1, 76]}.

When no LPF combination is found to fit within the roof of all operands then the smallest roof, identified as either the one with the smallest memory index or the one with the least space available, jumps to the next level available in the hierarchy. After that, a new search for the fitting combinations with the updated roof restarts.

Subsequent to each assignment is the placement of a virtual memory level separator, which creates a fictive memory level in which all operands have equal sets of LPFs, as in Figure \ref{fig:loop_blocking}. When all LPFs are assigned, the memory hierarchy is organized in virtual memory levels, within which its LPFs can be permuted. All the possible permutations are generated and sent for evaluation by the hardware cost model.

\begin{algorithm}[t]
\label{alg:bsg}
\caption{Blocking scheme generator}
\SetAlgoLined
\DontPrintSemicolon
Initialize EmptyScheme.roof and EmptyScheme.LPFList\;
Append EmptyScheme to PartialSchemes\;
\While{LPFList is not empty for all PartialSchemes}{
\For{Ps in PartialSchemes}{
\For{k in range(0, len(Ps.LPFList))}{
CombList $\gets$ combinations(Ps.LPFList, k)\;
\For{LPFComb in CombList}{
\If{fitComb(LPFComb, Ps.roof)}{
Update Ps with LPFComb\;
Remove LPFComb from LPFList\;
Append Ps with LPFList to PartialSchemes\;
}
}
}
}
}
\end{algorithm}

%\vspace{-1em}
\subsubsection{Shared and non-shared memory levels}
A shared memory level in the hierarchy is a level in which multiple operand types can be stored, and different operands can occupy different portions of the space available in the level depending on the blocking scheme assigned. The presence of shared memory levels greatly increases the amount of blocking scheme combinations since they act as levels which have a flexible upper bound for the space available for each operand: having fixed the minimum utilization rate of the shared level (usually 70\%), depending on the blockings already assigned, different operands can occupy a larger or smaller chunk of the storage space.

\subsection{Exhaustive Search}

Depending on the complexity of the hierarchy and the workload, this search method generates all possible schemes through loop blocking and loop reordering in an exhaustive way. It can take multiple hours to run the search and evaluate millions of valid mappings for a single layer, of which only a few are optimal ones. Speed-up techniques are required to explore the mapping space more efficiently, resulting in the heuristic and the iterative search introduced underneath.

\subsection{Heuristic Search Based On Data Reuse and Stationarity}
Once LPF assignment is completed, the data reuse for each operand and level can be extracted. If a particular combination of loop prime factors causes the data-reuse to be equal to 1 at a specific level in the hierarchy, it follows that that level is unnecessary since it causes useless memory accesses.

Consequently, the heuristic search discards all mappings with data reuse values equal to 1 for intermediate levels in the hierarchy (excluding the innermost and outermost ones). 
It is important to note that this rule does not hold for Input data, as even with data reuse equal to 1, this level may exhibit the FIFO effect and be optimal.

\begin{figure*}[h!]
\centering
\includegraphics[width=6in]{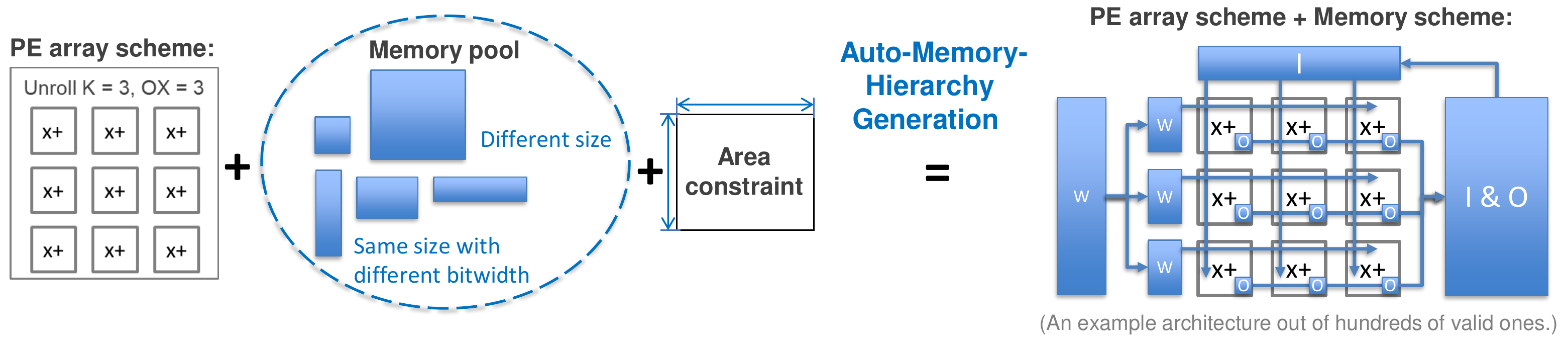}
% \vspace{-1.2em}
\caption{Memory hierarchy generator overview.}
\label{fig:msg}
\vspace{-1.2em}
\end{figure*}

\hl{Successive to this solution reduction step, the permutation of the LPFs is carried out again within the virtual memory levels to generate only those schedules that maximize stationarity for each operand (W/I/O), in order to avoid trying out all the permutations.} %\mv{How much do we save from this optimization? Do we report this somewhere?}
\subsection{Iterative Search Based On Early-Stage Cost Evaluation}
The last search strategy proposed explores the mapping space in an \textit{iterative} way, instead of generating an exhaustive list of blocking scheme combinations first and possibly pruning away the sub-optimal ones. Starting from the innermost level of the hierarchy, an iteration step consists in finding the set of LPFs that causes the largest amount of energy savings. After each iteration the best virtual memory level found is stacked upon those previously found ones.

Since this scheme analyzes partial schemes before converging to an optimal point and ignores the influence of the upper levels in the hierarchy when making a decision at the lower levels, it might reach a sub-optimal point in the temporal mapping space. We will show in the case studies later that the energy-overhead is however almost always within 5\%  of the cost of the optimal mapping. However, the speed-up with respect to the other methods (up to 10$\times$) and the much smaller memory footprint (\textasciitilde100s of mappings to be stored at each iteration step compared to the millions of the heuristic search) may be worth the trade-off for many DSEs.

% ----\input{52_arch_generator}------

\section{Architecture Generator}
\label{section_ag}
The performance of an embedded system is determined by the joint effect of the execution schedule combined with the hardware architecture, as observed in \cite{TIMELOOP}. The search for the optimal design should therefore not ignore the influence of the latter element, whose most influential component is the memory hierarchy.
 
% While the interconnection cost also plays a role, its energy component can be computed only by going through all the synthesis steps, (such as the case in \cite{MAGNET}) and can thus be hardly defined by a rapid analytical model.

The ZigZag architecture generator aims to autonomously generate all valid memory hierarchies %by finding the fitting combinations of memories from a pool 
in a search space 
constrained by area, PE array dimensions and spatial unrolling scheme. It draws the memories to build the hierarchy from a pool of available memory instances with different memory sizes and memory bandwidths. For each feasible memory hierarchy that fits the area constraint, the optimum memory bandwidth is selected based on the theory presented in Section \ref{sub_section_cost}. The addition of this generator effectively adds another dimension in the design space, % The presence of this module in the framework provides the designer with the possibility of identifying the optimal memory hierarchy and its optimal temporal mapping scheme, thus enabling the definition of 
yet, enables the designer to find the best point in the three-dimensional space described by area, energy consumption and throughput.

\subsection{Comprehensive Memory Pool Description}

Each memory in the pool is defined by its storage size expressed in bits, its cost of access expressed in pJ for read and for write, and its area in $\mu$m$^2$.

Given that the framework is also able to determine what would be the required memory bandwidth, so as to maximize the throughput of an architecture, each of the parameters of the memories in the pool are described for several distinct bandwidths as well.

The cost parameters (area, access energy for each bandwidth) have to be defined as input and are technology dependent: their accurate definition is vital to obtain the optimal design point. For running our estimation the CACTI 7.0 tool has been deployed at 65 nm, but these values can be fixed by the user before each simulation run so as to mirror the technology that is actually being deployed.
Each memory in the pool is characterized as in Table \ref{tab:mem_pool_sample}. 
\begin{table}[t]
\centering
\caption{Sample of 64 Byte RF in the pool for an $8\times8$ PE array. Tuple values = [ \textit{read}, \textit{write}] conditions.}
\begin{tabular}{|c|c|c|c|}
\hline
\textbf{Size [bit]}    & \multicolumn{3}{c|}{\small{2048}}                  \\ \hline
\textbf{Area [$\mu$m$^2$]}        & \small{4740.06}      & \small{4825.56}      & \small{7457.82}      \\ \hline
\textbf{Access cost [pJ]} & \small{[0.88, 0.98]} & \small{[0.99, 1.39]} & \small{[2.52, 3.49]} \\ \hline
\textbf{Mem bw [bit]}      & \small{[8, 8]}       & \small{[16, 16]}     & \small{[64, 64]}     \\ \hline
\textbf{Unroll}      & \multicolumn{3}{c|}{1, 8, 64}                     \\ \hline
\end{tabular}

% \vspace{-1.2em}
\label{tab:mem_pool_sample}
\end{table}

The unroll parameter specifies the amount of times the memory level is replicated in the architecture.
\subsection{Memory Hierarchy Generation}

Figure~\ref{fig:msg} gives an overview of the function of the memory hierarchy generator.

The generation of the set of valid memory schemes consists of three successive stages, respectively the \textit{fitting of the memories}, the \textit{operand assignment} and the \textit{bandwidth optimization}.

In the first stage the memory pool is firstly extended to include the unrolled version of the single memories as well, so as to have the possibility to have memory levels present in all PEs of the array or unrolled along a single dimension of the array. \hl{Subsequently all the fitting combinations with repetition (with max repetition set to 3 as the number of operands) of the memory elements from this enhanced pool are generated and assigned to an operand or to a set of operands as in the case of shared memory levels.}

The output of this stage is a list of valid memory hierarchies which are sequentially fed to the schedule generator, which in turn finds for each the optimal temporal mapping and its required bandwidth by means of the hardware cost model. When all hierarchies are analyzed, the optimal memory hierarchy and its optimal temporal mapping for a specific layer in a network will be identified.

% ----\input{6_validation}------

\begin{figure}[b]
    \begin{minipage}{.5\columnwidth}
		\centering
		\includegraphics[width=1.1\textwidth]{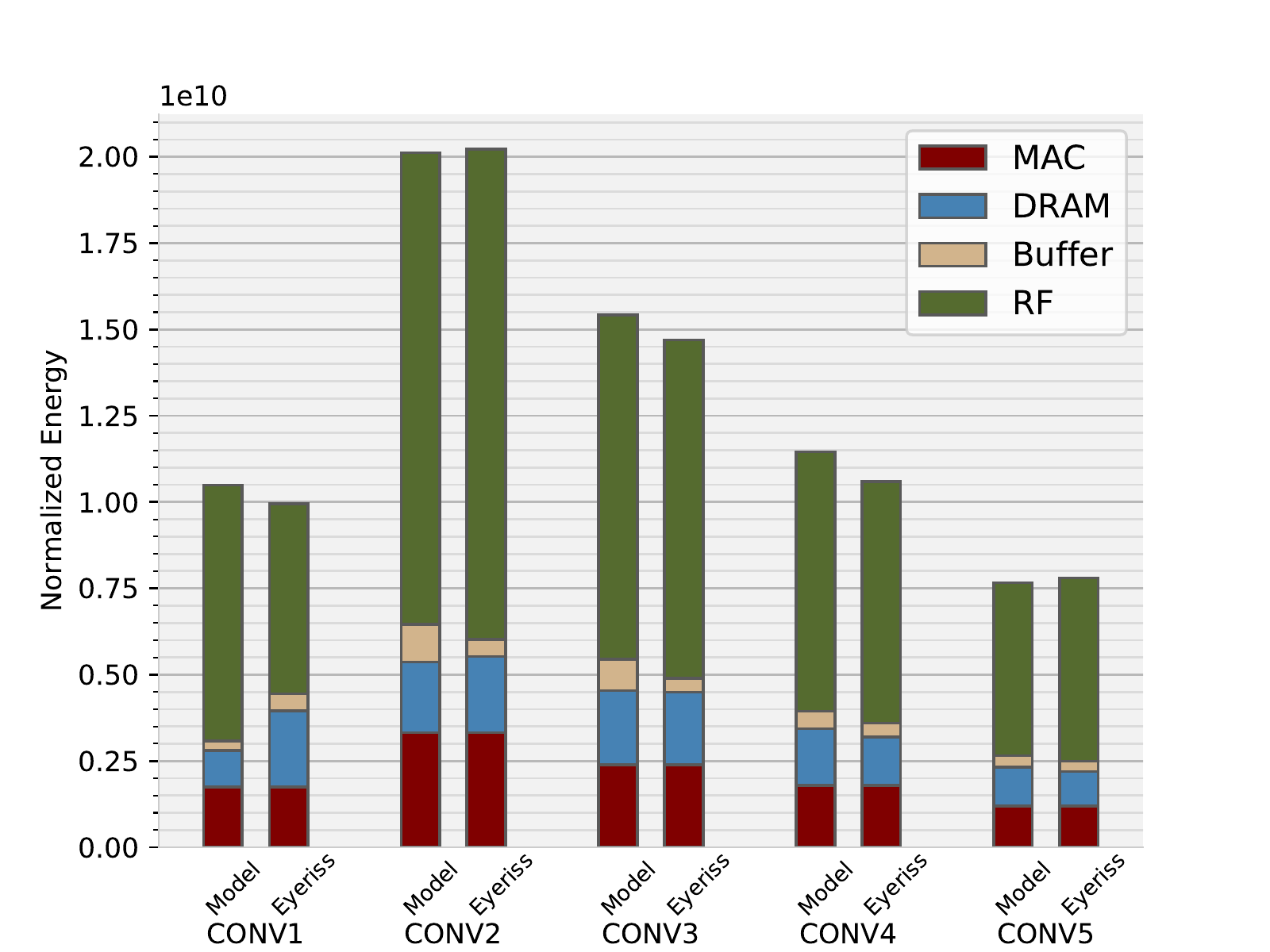}
	\end{minipage}%
	\begin{minipage}{.5\columnwidth}
		\centering
		\includegraphics[width=1.1\textwidth]{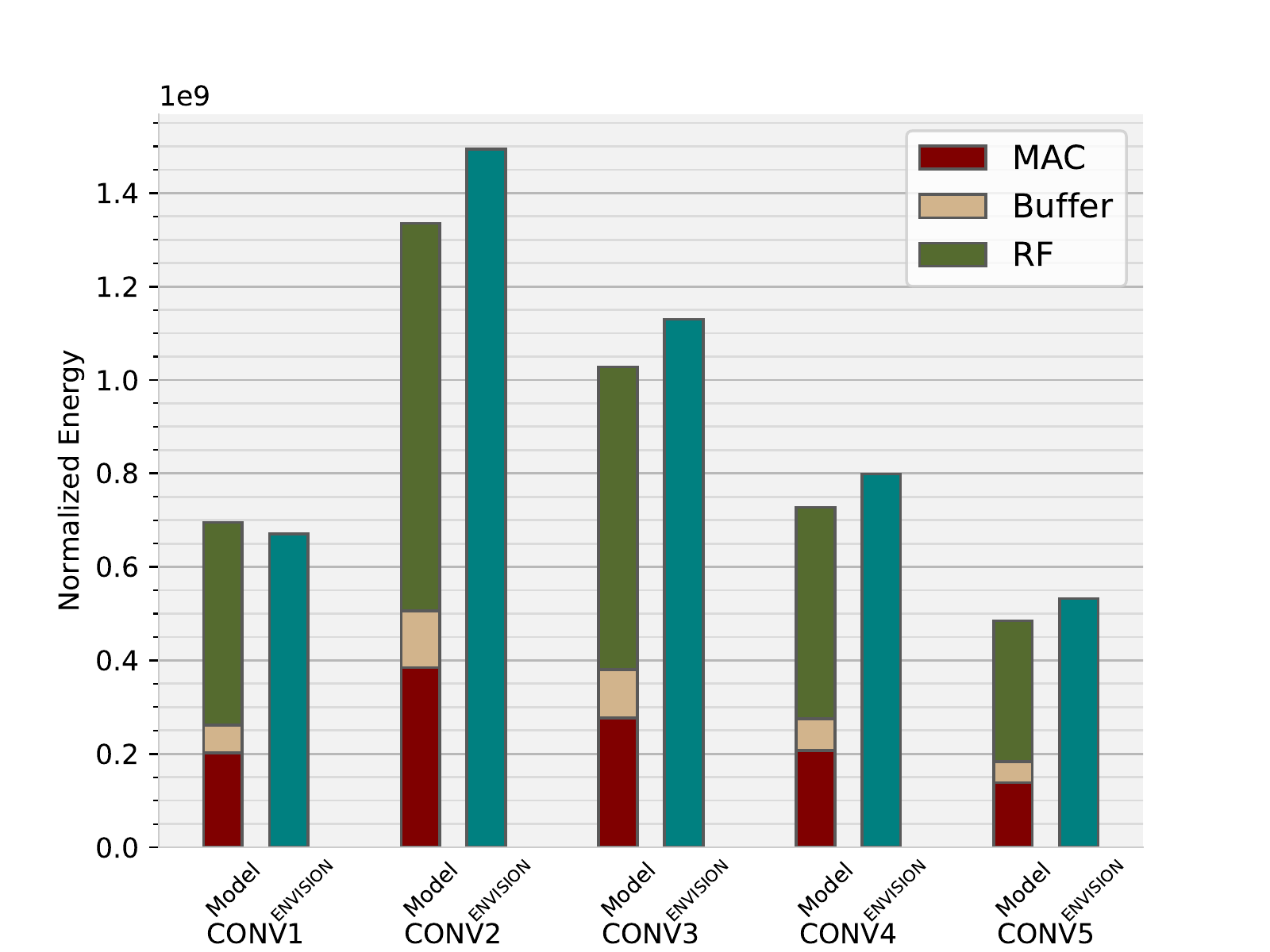}
	\end{minipage}
	%\vspace{-1.2em}
	\caption{Model validation of AlexNet CONV layers on Eyeriss (left) and ENVISION (right).}
	\label{validation}
% 	\vspace{-1.2em}
\end{figure}

\begin{figure}[b!]
    \begin{minipage}{.5\columnwidth}
		\centering
		\includegraphics[width=1.1\textwidth]{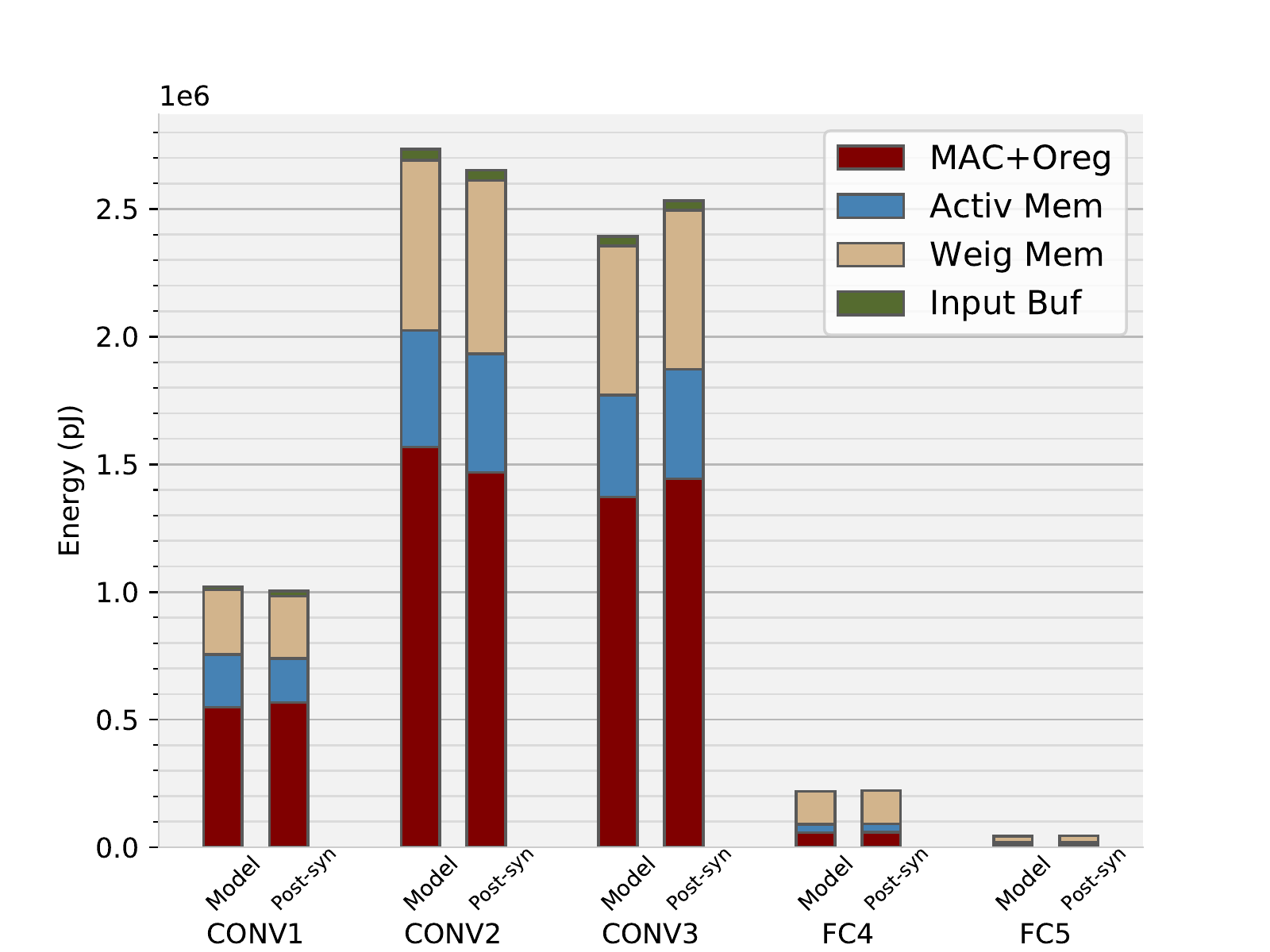}
	\end{minipage}%
	\begin{minipage}{.5\columnwidth}
		\centering
		\includegraphics[width=1.1\textwidth]{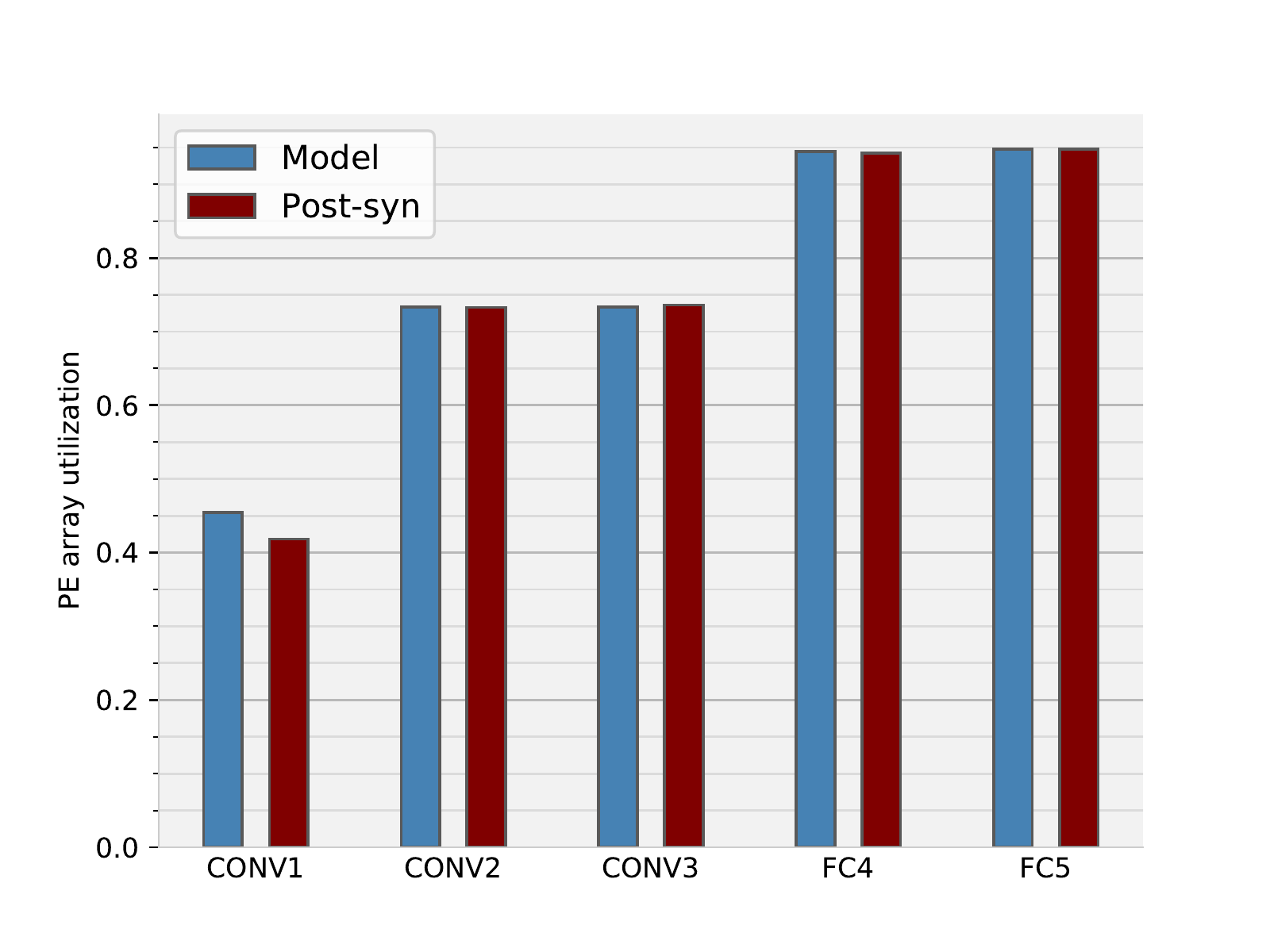}
	\end{minipage}
	%\vspace{-1.2em}
	\caption{\hl{Model validation against an in-house accelerator's post-synthesis results with a voice recognition workload on energy (left) and PE array utilization (right).}}
	\label{fig:validation2}
% 	\vspace{-1.2em}
\end{figure}

\begin{figure*}[h]
    \begin{minipage}{\columnwidth}
		\centering
		\includegraphics[width=0.98\textwidth]{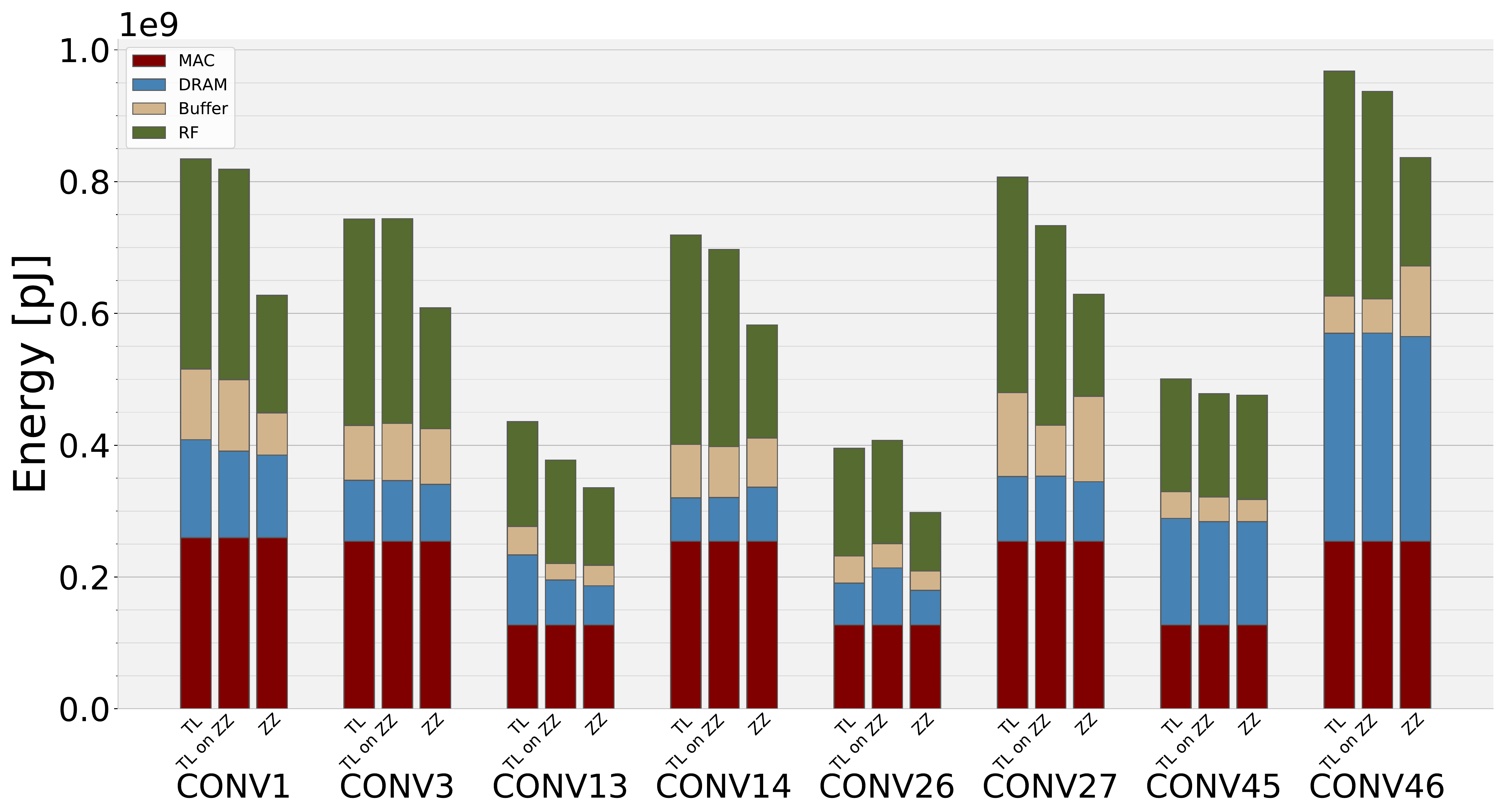}
	\end{minipage}%
	\begin{minipage}{\columnwidth}
		\centering
		\includegraphics[height=0.49\textwidth]{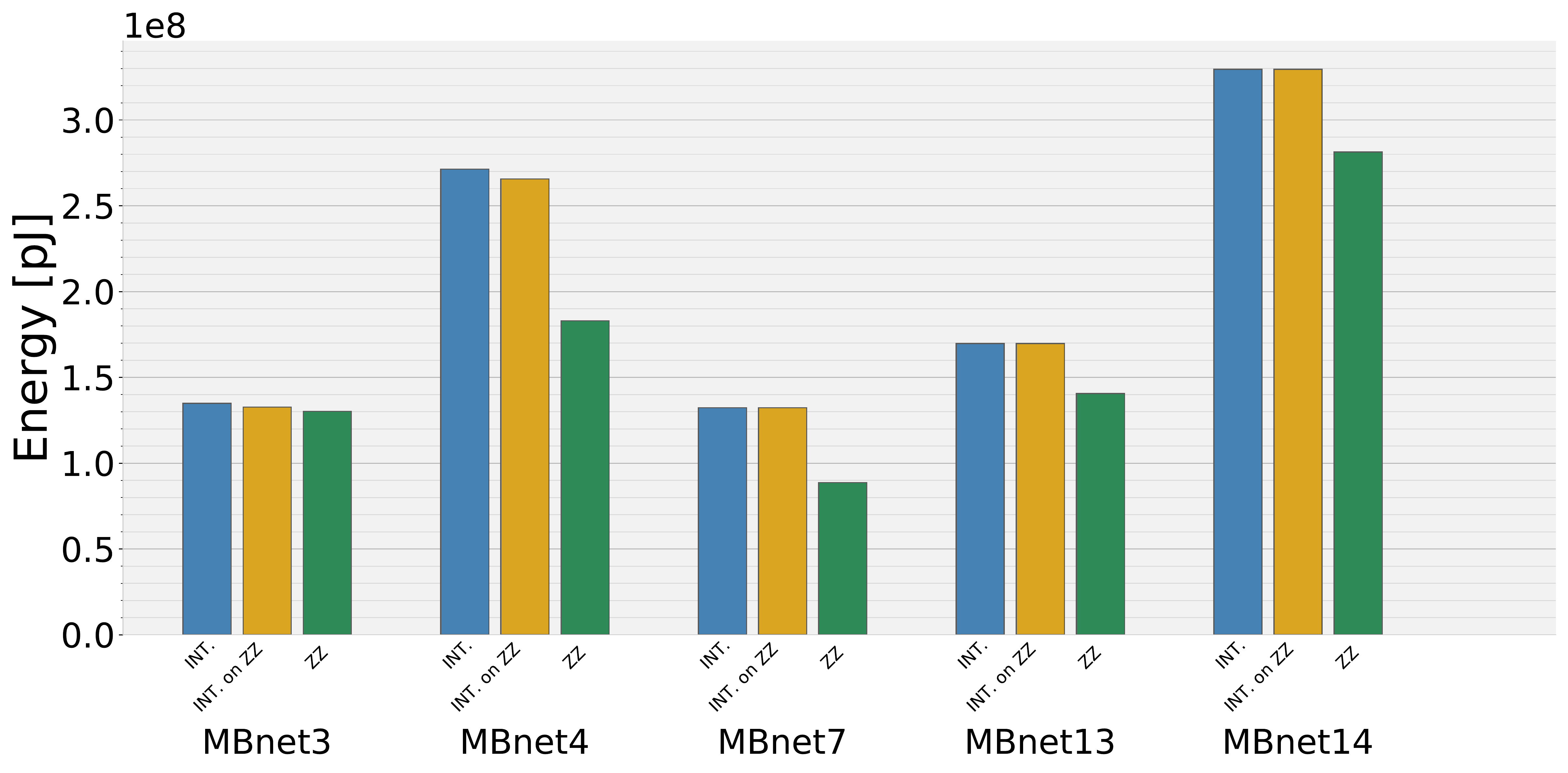}
	\end{minipage}
	%\vspace{-1.2em}
	\caption{Model validation against Timeloop+Accelergy (left) and Interstellar (right). In each group of three bars, the left/middle/right bars are respectively the best schedules found \& evaluated by SotAs, the best schedule found by SotAs \& evaluated by ZigZag, and the best schedule found \& evaluated by ZigZag.} 
	\label{fig:validation3}
% 	\vspace{-1.2em}
\end{figure*}

\section{Validation}
\label{section_validation}
%The cost model is validated against the Eyeriss and ENVISION architectures, by comparing the framework estimated cost with the reported ones. For the Eyeriss case, the dataflow and the memory hierarchy parameters adopted are those present in \cite{EyerissV1} whereas for ENVISION the dataflow and hierarchy are those reported in \cite{ENVISION} and the normalized energy figures correspond to the values at full precision without voltage scaling and sparsity reduction. For the former the estimations are within an acceptable 5\% margin error while for the latter the error margin are on average 7.5\%.

The hardware cost model and the mapping engine are validated with three methodologies: a.) against published taped-out chips measured results; b.) against in-house post-synthesis extracted energy and performance data; c.) against other DNN accelerator DSE frameworks.

Firstly, we model the dataflow and hardware architectures of both Eyeriss \cite{EyerissV1} and ENVISION \cite{ENVISION} and compare the estimated energy \hl{(left bars)} with their reported values \hl{(right bars)}, as depicted in Figure \ref{validation}. %The dataflow and memory hierarchy configuration for Eyeriss and ENVISION adopted during this validation are borrowed from \cite{EyerissV1} and \cite{ENVISION}. 
The resulting energy values, \hl{normalized with respect to a single MAC cost,} are shown for full precision operation without voltage scaling or sparsity reduction. The estimated values are within an acceptable 5\%, resp. 7.5\% error margin. % error for Eyeriss and for ENVISION the observed average error margin is 7.5\%.  
%\mv{not clear in the figure what is the estimation and what the number from the paper}\ph{IF WE STILL HAVE SPACE DEFINITELY}

Secondly, the validation of the energy as well as the performance model is performed against a complete in-house accelerator at RTL level, shown in Figure~\ref{fig:validation2}. A maximum error of 6\% of energy and 9\% of PE array utilization are achieved.

Finally, the validation of the cost model as well as the temporal mapping generator is carried out against two SotA frameworks: Timeloop \cite{TIMELOOP} + Accelergy \cite{Accelergy}, resp. Interstellar \cite{Yang2018}, as shown in Figure~\ref{fig:validation3}. For Timeloop + Accelergy, the ResNet34 \cite{ResNet} convolutional layers are first mapped on the Eyeriss hardware architecture. Subsequently we estimate the energy cost of this (suboptimal) mapping through Timeloop + Accelergy (`TL' in Figure~\ref{fig:validation3} left), and ZigZag (`TL on ZZ'), matching within 10\%. Yet, when we let our temporal mapping generator optimize the scheduling for the same architecture and workload, more optimal design points are found, which can lead to up to 20\% energy savings (`ZZ'). The optimal mapping exploits uneven mapping, and can not be validated back with Timeloop, since it cannot be represented by Timeloop's limited design representation. Note that a lot of ReNet34 layers have the same dimension size, thus we only pick all the non-repetitive layers for validation.

%The same validation method is applied to Interstellar; however in this case, even though the two cost models match, the optimal mapping found by ZigZag does not lead to significant improvements.}\hldelete{still to be confirmed with other hw templates...} \mv{the figure shows significant imprivements no?}  \mv{I would remove Interstellar. Both for space reasons, and because we had to delete all kinds of components...?  Everyone trusts Accelergy the most anyway, and no-one trusts Interstellar...} \mv{Instead, merge fig 13 and fig 14. E.g. fig 13 can put energy on the left Y-axis, and throughput on the right Y-axis (just a star on top of each energy bar), this becomes 13.a; and Timeloop comp then becomes 13.b.  This saves space, which you can then use to explain fig 13.a...\\ Be careful... In case you decide to drop Interstellar, we have to slightly adapt the rebuttal letter.}
A similar validation method is applied to Interstellar on a hardware template with three  all-shared memory levels. Several pointwise layers in MobileNet V1 \cite{MobileNets} are used for testing, since their framework cannot handle the `pr' loop pair data reuse accurately. Note that in this two-step experiment (firstly matching energy with Interstellar's best schedule, and then searching for an better one in our enlarged design space), we ignored the same energy contributions that Interstellar ignored, such as distinguishing psums and final sums, separating cost of memory writing from reading, etc. The result shows that an total energy matching with only 3\% error is achieved, and our uneven mapping scheme outperforms its best even mapping scheme by up to 33\% concerning energy.

% ----\input{7_case_study}------

\section{Case Studies}
\label{section_case_studie}

To better understand the vast design space and show the strength of ZigZag, three case studies from different design abstraction levels are conducted.

%\subsection{Case Study Settings}
\subsection{Case Study 1: Impact of Scheduling}

\begin{figure}[t]
    \centering
    \includegraphics[width=3.3in]{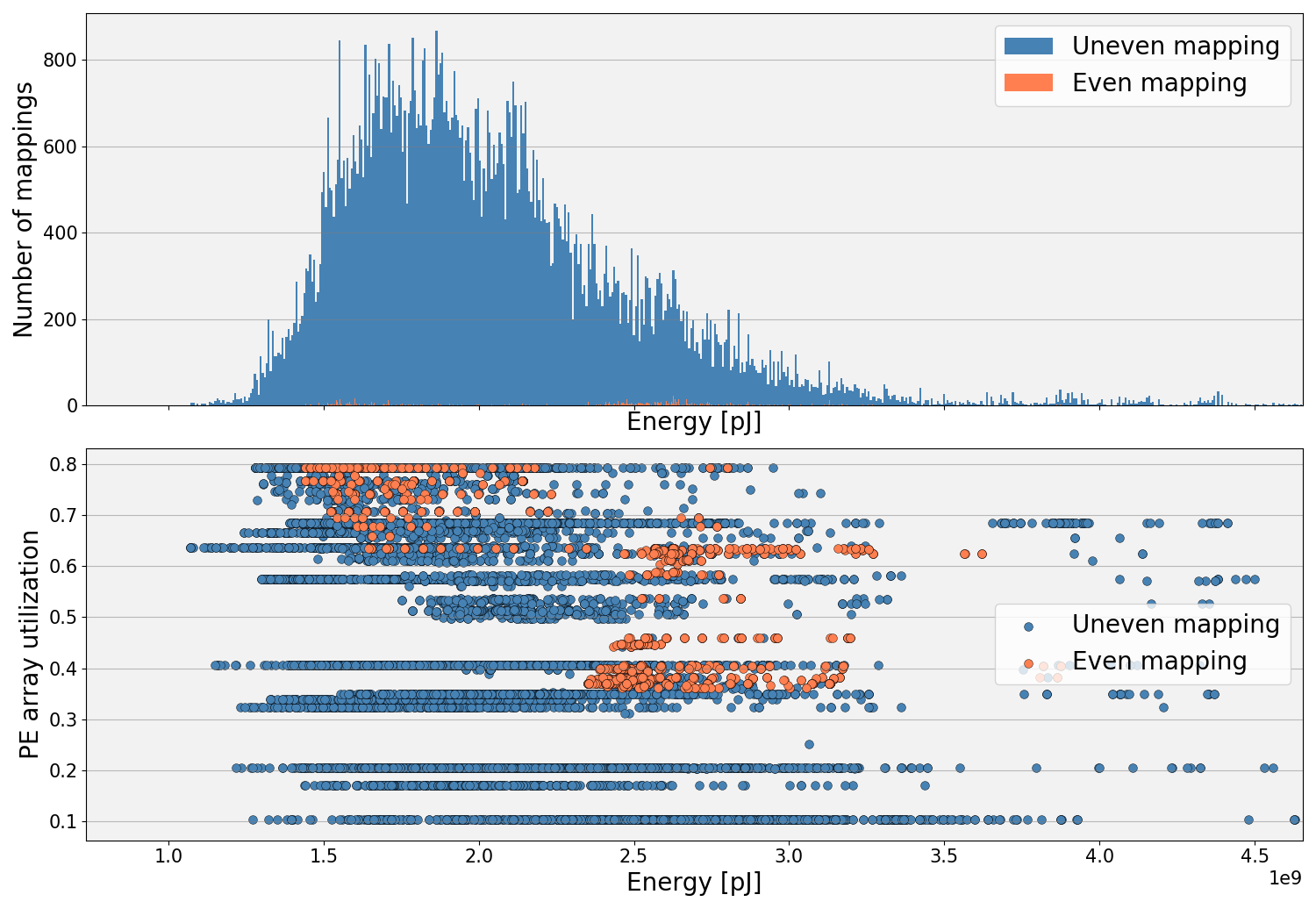}
    % \vspace{-1.2em}
    \caption{\hl{Schedule's impact on energy and throughput with even/uneven blocking on AlexNet CONV2 mapped on Eyeriss (access energies derived from CACTI7).}}
    % \vspace{-0.6em}
    \label{fig:cs1_conv2_hist}
\end{figure}

The cost estimator and temporal mapping generator of ZigZag are used to assess the impact of scheduling on both energy and throughput. This is assessed for AlexNet convolutional layer 2 on an Eyeriss-like architecture. \hl{A memory bandwidth of 16 bit/cycle is assumed for RF and 64 bit/cycle for GLB.} The results in Figure~\ref{fig:cs1_conv2_hist} shows there is an up to \hl{$4.7\times$} energy variance and an up to \hl{$8\times$} throughput variance across temporal schedules.

\hl{A striking observation that can be made is how limited the space of exploration is if only even mappings are considered: the number of uneven mappings is thousands of times larger than the number of the even ones and the uneven ones can reach optimal design points that would be otherwise not achievable. For this particular case study an improvement of 25\% of the energy value can be obtained with respect to the best even mapping. Similarly, as is the case for the validation tests run in the previous section, comparable improvements are achieved for different architectures and workloads as well.}
% \vspace{-0.6em}

\begin{figure}[t]
    \centering
    \includegraphics[width=3.3in]{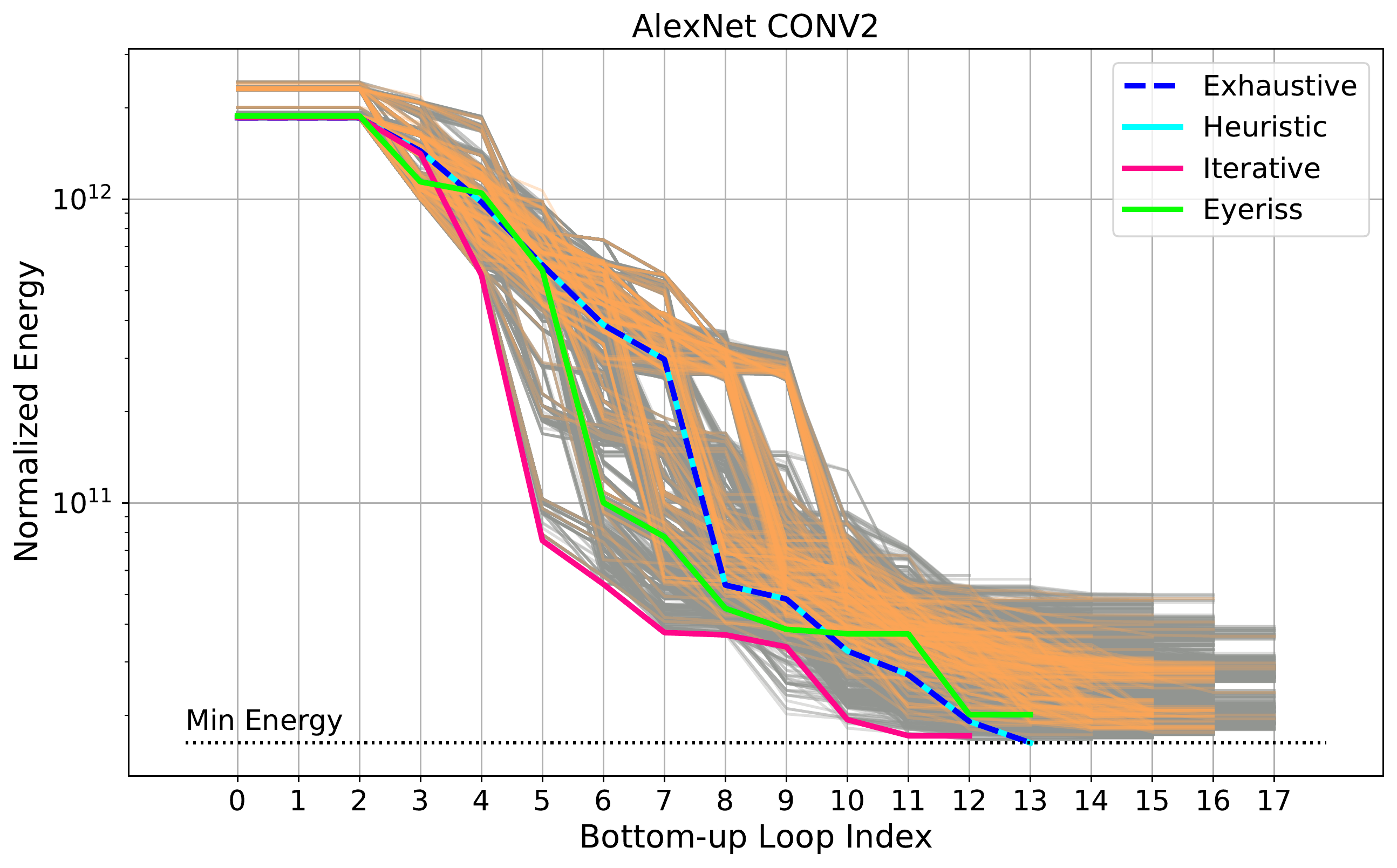}
    % \vspace{-1.2em}
    \caption{Visualization of individual loop’s impact on energy saving and three searching strategies' trajectory.}
    % \vspace{-1em}
    \label{fig:cs1_Multi_schedule_energy}
\end{figure}

Next, the three ZigZag search engines (Section 
\ref{section_tmg}) are compared in terms of their searching efficiency. Figure~\ref{fig:cs1_Multi_schedule_energy} visualizes their search procedures and obtained results. The figure contains one line for each represented schedule. For each schedule, this line depicts the energy spent on memory accesses assuming all lower level loops are scheduled, and all upper level loops are assigned to DRAM (Figure~\ref{fig:loop_vis}a converted to energy). The rightmost point of every curve is the actual energy of the completely scheduled workload.  
%These energy savings are relative to the energy when all data access  Figure~\ref{fig:loop_vis}a we see that from bottom level (MAC) to top level (DRAM), each ir loop contributes to data access saving. Eventually, these data access savings will turn into energy savings, which Figure~\ref{fig:cs1_Multi_schedule_energy} captures. 

In Figure~\ref{fig:cs1_Multi_schedule_energy}, the grey curves are thousands of randomly-sampled valid schedules the tool found in exhaustive search, while the orange curves give results from the heuristic search. The iterative search only has 1 trajectory, as it only refines a single solution. The bold curves mark the trajectory of the minimal-energy schedules found by each strategy, plus the schedule reported in Eyeriss paper. Notice that the best schedule found by exhaustive search and heuristic search overlap, meaning that the heuristic search can equally well locate the global optimum schedule as the exhaustive search does. Iterative search resulted in another schedule, which is slightly (5.5\%) more energy consuming than the global optimum. The Eyeriss schedule is 23.8\% worse than the global optimum.
Table~\ref{table_search_comp} gives an overall comparison of using these three searching strategies on locating the best schedule for AlexNet conv. layer 1-5. It shows that heuristic search can bring a $2.5\times$ speedup without losing optimality, while iterative search brings a $7.5\times$ speedup with a 1.6\% energy penalty on average.

% \vspace{-0.8em}

\begin{table}[t]
\centering
\caption{Comparison on three searching strategies}
\label{table_search_comp}
\scalebox{1}{
\begin{tabular}{|l|c|c|c|}
\hline
\multicolumn{1}{|c|}{AlexNet CONV1-5} & \makecell{\textbf{Exhaustive} \\ \textbf{Search}} & \makecell{\textbf{Heuristic} \\ \textbf{Search}} & \makecell{\textbf{Iterative} \\ \textbf{Search}} \\ \hline
\begin{tabular}[c]{@{}l@{}} Total Number of\\ Valid Schedules\end{tabular} & 4887334 & 1444608 & \begin{tabular}[c]{@{}c@{}}Partial: 3810973\\ Final: 7168\end{tabular} \\ \hline
\begin{tabular}[c]{@{}l@{}}Relative Number \\ of Schedules\end{tabular} & 100\% & 30\% & \begin{tabular}[c]{@{}c@{}}Partial: 78\%\\ Final: 0.15\%\end{tabular} \\ \hline
Elapsed Time (sec) & 13228.2 & 5330.4 & 1760.1 \\ \hline
Speedup & $\times$1 & $\times$2.5 & $\times$7.5 \\ \hline
Minimum Energy & 1 & 1 & 1.016 \\ \hline
\end{tabular}
}
\end{table}
% \vspace{-0.8em}

\begin{table}[t]
\centering
\small
\caption{Case study 2 input parameters}
\begin{tabular}{|c|p{5cm}|}
\hline
\textbf{PE array size}     & 8 $\times$ 8  \\ \hline
\textbf{Spatial unrolling} & OX | K \\ \hline
\textbf{Memory pool} & 64 Byte, 1KByte, 4KByte, 16KByte, 128 KByte, 2 MByte @ 65nm \\ \hline
\textbf{Workloads} & MobileNet V2 - L13, L15, L39, L46  \\ \hline
\textbf{Temporal mapping} & Optimal with heuristic search \\ \hline
\end{tabular}
\label{tab:cs2_input}
\end{table}
%\vspace{-1.4em}

\subsection{Case Study 2: Workload and Memory Hierarchy}

%\seb{comment: repetition of the term optimal. Maybe try to rephrase the next text since the design is not cooptimizing, it is the framework, right?}
The best design co-optimizes the dataflow schedule and the hardware architecture. Here, every workload (e.g. neural network layer) would have a different optimal memory hierarchy and dataflow schedule.

%This implies that different workloads (e.g. layers in a network) should have different memory hierarchies to achieve the maximum performance, which 
Yet, in reality, in most designs this is impossible as the memory levels are hardwired on chip. This raises the question of whether the flexibility overhead from reconfigurable memories with a network-on-chip (able to dynamically change the memory hierarchy between layers) is amortized by its benefits.
%This raises the question whether it is worth the flexibility overhead to have reconfigurable memories with a network-on-chip, that could dynamically change the memory hierarchy between layers. %This would mean that the memory hierarchy changes from layer to layer by commuting the type of operands that are stored in the memory levels on the chip.
To evaluate this, we deployed the architecture generator and the temporal mapping generator on different layers from MobileNetV2, namely the layers 13, 15, 39 and 46. Each layer is constrained to the same PE array with the same spatial unrolling, and the memory hierarchy and temporal mapping are jointly-optimized for minimal energy consumption. The input parameters are listed in Table \ref{tab:cs2_input}.

\begin{table*}[t!]
\centering
\caption{Estimated energy for different workloads and their optimal architectures.}
\scalebox{0.9}{
\begin{tabular}{ccccccc}
& \multicolumn{1}{c}{Optimal memory hierarchy and its spatial unrolling}  & Run L13 & Run L15 & Run L39 & Run L46 & \textbf{Total} \\ \cline{2-7}

\multicolumn{1}{l|}{L13 Arc.} & \multicolumn{1}{l|}{\makecell{\tiny{\texttt{{W: [16777216], I: [32768, 16777216], O: [512, 32768, 16777216]}}} \\ \tiny{\texttt{W: [[(K, 8), (OX, 7)],[]], I: [[(OX, 7)], [(K, 8)],[]], O: [[],[(OX, 7)],[(K, 8)],[]]}}}} & \multicolumn{1}{c|}{\cellcolor[HTML]{00FF00}{\color[HTML]{000000} \small{27.57} $\mu$J}} &  \multicolumn{1}{c|}{\small{60.71 $\mu$J}}  & \multicolumn{1}{c|}{\small{59.47} $\mu$J} & \multicolumn{1}{c|}{\small{44.79} $\mu$J} & \multicolumn{1}{c|}{\small{192.54} $\mu$J} \\ \cline{2-7} 

\multicolumn{1}{l|}{L15 Arc.} & \multicolumn{1}{l|}{\makecell{\tiny{\texttt{{W: [16777216], I: [512, 8192, 16777216], O: [512, 32768, 16777216]}}} \\ \tiny{\texttt{W: [[(K, 8), (OX, 7)],[]], I: [[(OX, 7)], [(K, 8)],[]], O: [[],[(OX, 7)],[(K, 8)],[]]}}}}  & \multicolumn{1}{c|}{\small{58.79 $\mu$J}} & \multicolumn{1}{c|}{\cellcolor[HTML]{00FF00}{\color[HTML]{000000} \small{28.37} $\mu$J}} & \multicolumn{1}{c|}{\small{60.22} $\mu$J} & \multicolumn{1}{c|}{\small{41.69} $\mu$J} & \multicolumn{1}{c|}{\small{189.07} $\mu$J} \\ \cline{2-7}

\multicolumn{1}{l|}{L39 Arc.} & \multicolumn{1}{l|}{\makecell{\tiny\texttt{W: [32768, 16777216], I: [16777216], O: [512, 32768, 16777216]} \\ \tiny\texttt{W: [[(K, 8), (OX, 7)], [], []], I: [[(K, 8), (OX, 7)], []], O: [[], [(K, 8)], [(OX, 7)], []]}}}  & \multicolumn{1}{c|}{\small{41.69 $\mu$J}} & \multicolumn{1}{c|}{\small{40.31 $\mu$J}} &  \multicolumn{1}{c|}{\cellcolor[HTML]{00FF00}\small{49.51 $\mu$J}} & \multicolumn{1}{c|}{\small{45.09 $\mu$J}} & \multicolumn{1}{c|}{\small{176.6} $\mu$J} \\ \cline{2-7}

\multicolumn{1}{l|}{L46 Arc.} & \multicolumn{1}{l|}{\makecell{\tiny\texttt{W: [16777216], I: [32768, 16777216], O: [512, 32768, 16777216]} \\ \tiny\texttt{W: [[(K, 8), (OX, 7)], []], I: [[(K, 8), (OX, 7)], [], []], O: [[], [(OX, 7)], [(K, 8)], []]}}} & \multicolumn{1}{c|}{\small{29.25 $\mu$J}} & \multicolumn{1}{c|}{\small{64.15 $\mu$J}} & \multicolumn{1}{c|}{\small{51.34 $\mu$J}} &
\multicolumn{1}{c|}{\cellcolor[HTML]{00FF00}\small{34.37 $\mu$J}}  & \multicolumn{1}{c|}{\small{179.11$\mu$J}}  \\ \cline{2-7} 
 & \multicolumn{5}{|c|}{\textbf{Flexible architecture}} & 
    \multicolumn{1}{c|}{\cellcolor[HTML]{00FF00}{139.82 $\mu$J}}  \\  \cline{2-7}
\end{tabular}
}
\label{tab:CS2_STUDY}
\end{table*}

\begin{figure}[t]
    \centering
    \includegraphics[width=3.3in]{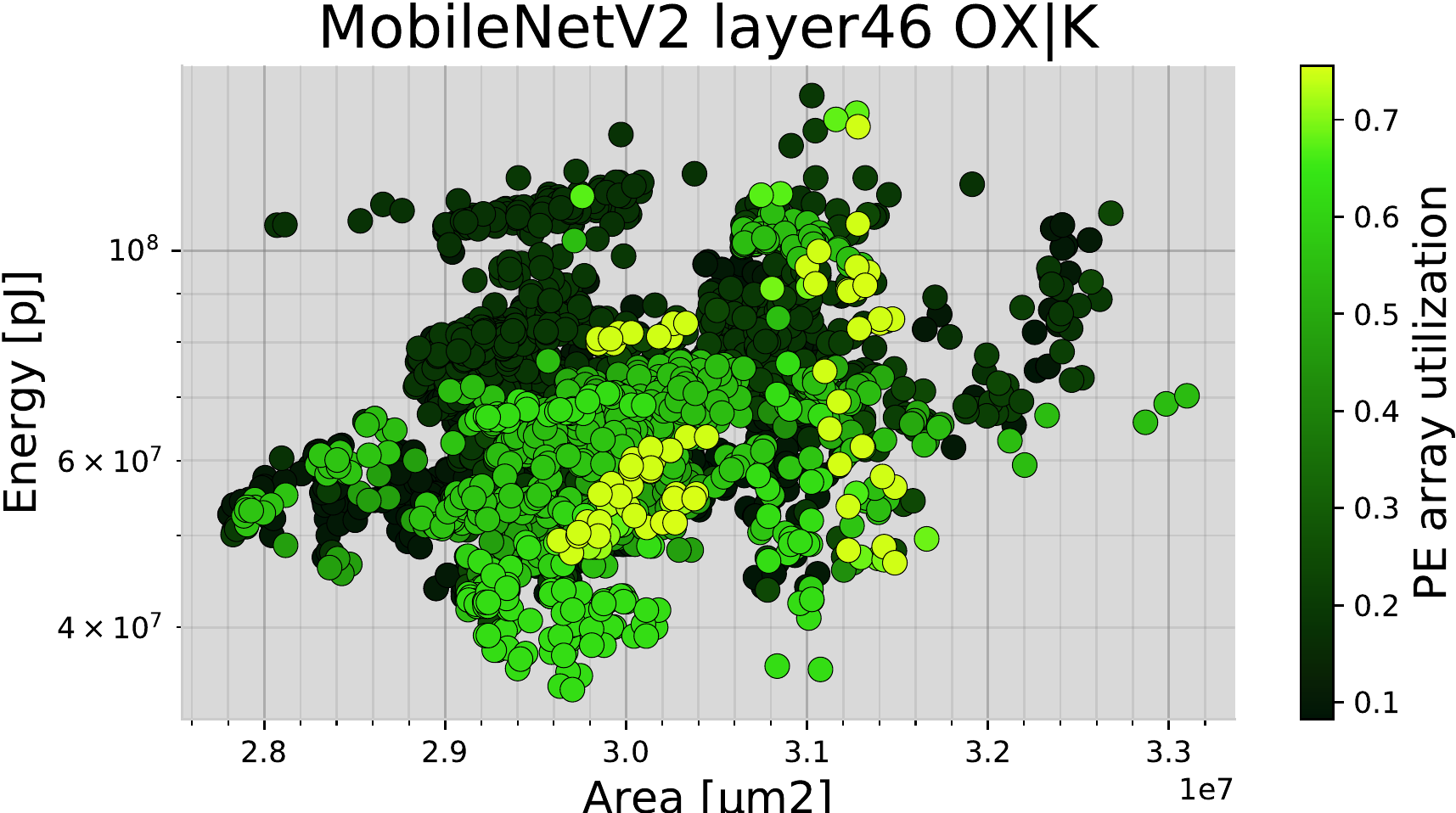}
    %\vspace{-1.2em}
    \caption{Design points identified by the framework for L46 of MobileNetV2. Each dot corresponds to a different memory hierarchy solution.}
    %\vspace{-0.8em}
    \label{fig:cs2_MN46}
\end{figure}

Table \ref{tab:CS2_STUDY} summarizes the result of this study. It suggests that having a flexible hierarchy may be worth the trade-off if the energy cost overhead of having a Network-on-Chip is within the 30\% of the total inference cost of the layer.
Figure~\ref{fig:cs2_MN46} visualizes the design space targeting on one single layer, showing the energy-performance-area tradeoff between hundreds of valid design points.

\subsection{Case Study 3: Spatial Unrolling and Memory Hierarchy}

\begin{figure*}[t]
	\centering
	\begin{minipage}{1\columnwidth}
		\centering
		\includegraphics[width=1.1\textwidth]{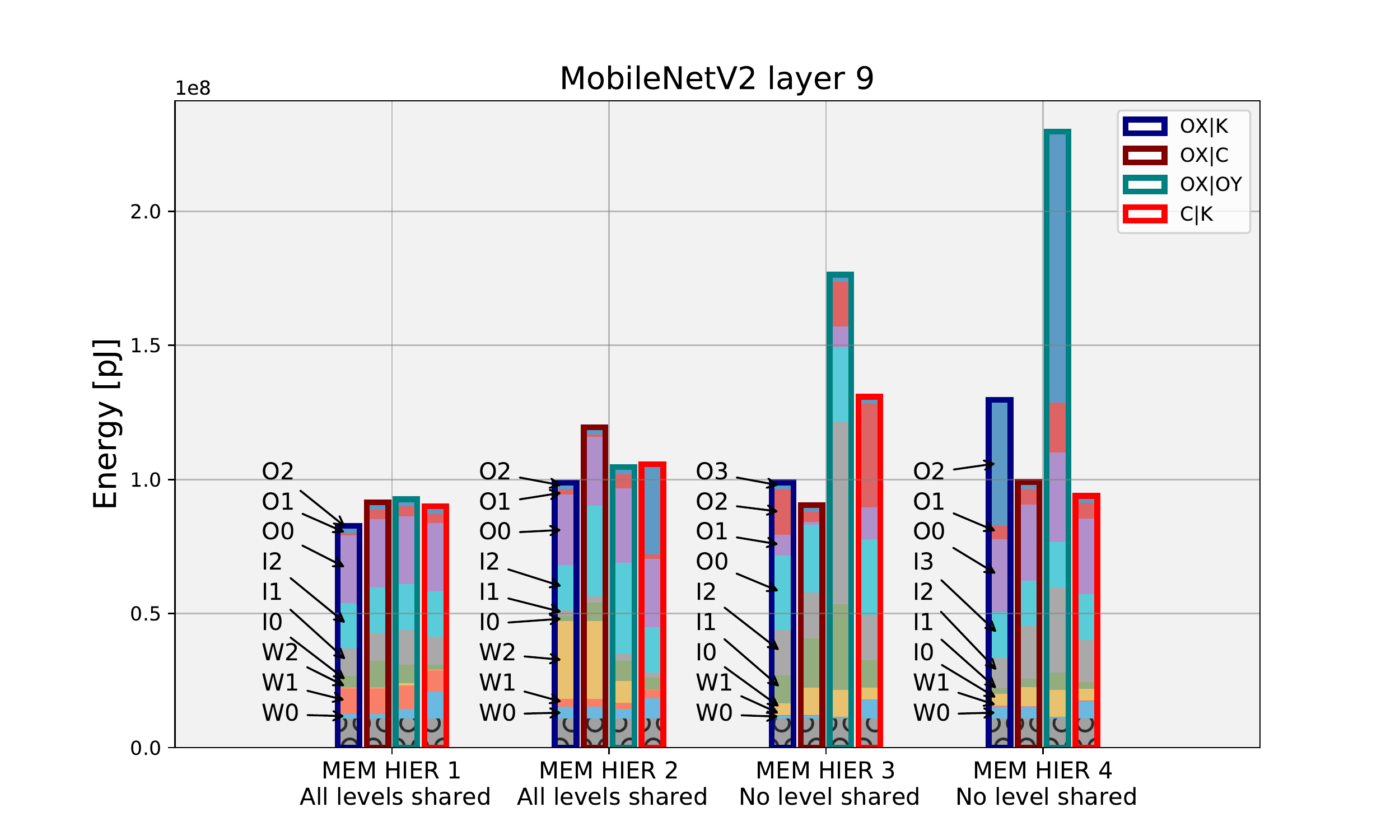}
		%\caption{MobileNet V2 - layer 9}
% 		\label{label1}
	\end{minipage}%
	\begin{minipage}{1\columnwidth}
		\centering
		\includegraphics[width=1.1\textwidth]{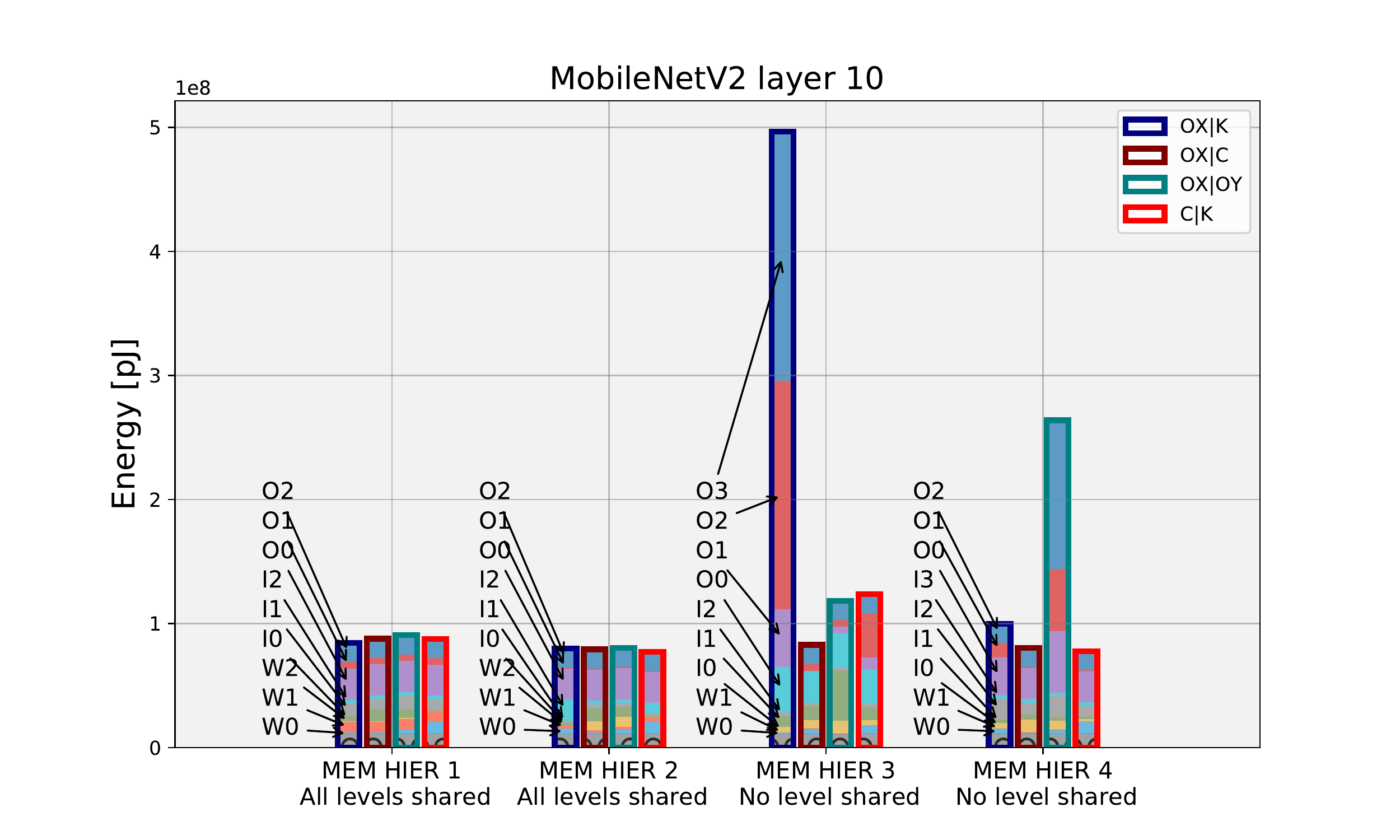}
		%\caption{MobileNet V2 - layer 10}
% 		\label{label2}
	\end{minipage}
	\caption{Influence of spatial unrolling on Shared vs Not-shared memory hierarchies.}
	\label{cs3}
\end{figure*}

Another degree of freedom in the design space to assess is the PE array's spatial unrolling. Previous works \cite{Yang2018} stated that the spatial unrolling has a very limited effect on the energy consumption as long as the PE array is fully mapped.
%The architectural design choice of which spatial unrolling to support also  may have a diverse impact on the cost depending on the type of memory hierarchy that is deployed.

Figure \ref{cs3} shows the energy consumption of two different layers of MobileNetV2 for several spatial unrollings and memory hierarchies. The number after operands (W/I/O) indicates memory level, e.g. W0 means the 0th (innermost) memory level (usually register file) of Weight. The leftmost two memory hierarchies have all their levels shared among the operands, the rightmost two are hierarchies with different memory levels for each operand. The results indicate that spatial unrolling has limited energy impact only when memory levels are shared between all operands.
%It is quite evident how the fact of sharing all memory levels may lead to the aforementioned compensation: while the last two hierarchies have widely distributed energy costs, the first two with shared memories have an overall cost that is practically identical for the different unrolling schemes. 
% This intuitively stems from the fact that a hierarchy with only shared memory levels compensates for the inefficiencies of a sub-optimal spatial unrolling by acting as a memory scheme with flexible level boundaries that enable a much larger number of temporal mappings.
It is because having shared memories softens the constraint on the memory utilization and makes the occupying size of each operand (W/I/O) at each memory level flexible, thus enabling a much larger number of temporal mappings, making it possible for the temporal mapping to adapt itself to the spatial unrolling. In other words, temporal mapping can compensate the unbalanced data reuse distribution in spatial unrolling among W/I/O operands with all-shared memory hierarchies.

Yet, the rightmost two hierarchies have no memory sharing among the different operands and have widely distributed energy costs: in this scenario such compensation is not possible, as the Temporal Mapping Generator has less degrees of freedom to arrange the schedule and balance the temporal and spatial data reuse. The spatial unrolling scheme here impacts energy efficiency up to 5$\times$.  
%\vspace{-0.4cm}

% ----\input{8_conclusion}------
% \vspace{-0.2cm}
\section{Conclusion}
\label{conclusion}
This paper presents ZigZag, a memory-centric rapid design space exploration framework for DNN accelerators. 

%Three modules work together to enable  the exploration of a much larger space of solutions with respect to the SotA.
Three modules cooperate in synergy to enable the exploration of a much broader space of solutions with respect to the SotAs.
%Firstly, the Architecture Generator is capable of generating all possible memory hierarchies given a PE array spatial unrolling scheme, a memory pool, and a area budget.
Firstly, the Architecture Generator is capable of generating all valid memory hierarchies (balanced/unbalanced, shared/separate) given a set of high-level hardware constraints.
%Secondly, with innovative searching strategies, Temporal Mapping Generator can effectively locate the optimal schedule for any type of memory system generated by Architecture Generator, optimizing either for energy or latency. 
Secondly the Temporal Mapping Generator can rapidly locate the optimal schedule (even/uneven) by means of innovative searching methods for any type of memory hierarchy provided by the Architecture Generator.
%Thirdly, with the memory-centric dataflow representation and the Loop Relevance Principle, Hardware Cost Estimator can easily predict energy and throughput for the highly flexible dataflows generated by Temporal Mapping Generator.
Thirdly, with the memory-centric dataflow representation and the Loop Relevance Principle, the Hardware Cost Estimator can analytically calculate energy and throughput for the schedules generated by the Temporal Mapping Generator.

Three case studies disclose the vast DNN accelerator design space from different perspectives. The first experiment shows the importance of adopting an optimal schedule when mapping the algorithm onto the hardware since it has huge impact on both energy and performance and uneven mappings opens up the searching space and leads to find better design points. The second experiment highlights that different workloads lead each to their own optimum memory hierarchies; it also assesses whether a Network-on-Chip that enables configurable memory bypassing and memory operand re-assigning is worth the implementation so as to enable the mapping of each workload on its own optimal memory scheme.
The third experiment partly disproves the conclusion drawn by Yang. et al~\cite{Yang2018} that spatial unrolling is unimportant as long as the PE array is fully mapped. Our results shows that this conclusion can only hold for memory hierarchies with all shared levels, in which the operand with less spatial data reuse can be well compensated by its temporal data reuse in local storage; in  memory hierarchies with not all the levels shared, the choice of the spatial unrolling can instead greatly affect the overall cost.

In conclusion, we showed the great capabilities and the uniqueness of ZigZag in exploring the design space of DNN accelerator. 

The research team is continuing building and polishing ZigZag. At the same time, we have open-sourced this project at \href{https://github.com/ZigZag-Project/zigzag}{https://github.com/ZigZag-Project/zigzag} and welcome comments and contributions from the community.

%In conclusion, we showed the great capabilities and potential of ZigZag in exploring the design space for DNN accelerators.

%\mv{I would like to read the conclusion too before you submit.}

%\mv{We have to organize the page layout such that everything (except the references) is max 11 pages, and the references appear on the 12th page.\\
%A MICRO paper can have as many references as you want (this does not count for the page count), and it is seen as a good thing to have many references. So if you can think of more (even slightly related) papers, it would be good to add some more references. Maybe e.g. add at least 1-2 additional references from our own team, as well as some extra for other people's work.}

% Add ACKNOWLEDGMENTS later
\section*{ACKNOWLEDGMENTS}
This research received funding from the Flemish Government (AI Research Program) and the Fund For Scientific Research Flanders (FWO-Vlaanderen).
%%%%%%% -- PAPER CONTENT ENDS -- %%%%%%%%

%%%%%%%%% -- BIB STYLE AND FILE -- %%%%%%%%

\bibliographystyle{IEEEtran}
\bibliography{refs}
%%%%%%%%%%%%%%%%%%%%%%%%%%%%%%%%%%%%

% that's all folks
\end{document}